\documentclass[journal]{IEEEtran}
\usepackage{epsfig,rotating,setspace,latexsym,amsmath,epsf,amssymb,amsfonts,bm,theorem,epstopdf,cite,mathtools,caption,subcaption,enumerate,longtable,accents,bbm,algorithm,algorithmic,graphicx,epsf,authblk,url,color,multirow,comment,tabularx,dsfont,physics,tikz,dsfont,booktabs}
\usepackage[inline]{enumitem}

\newcommand{\E}{\mathbb{E}}

\newcommand{\ind}[1]{\mathbf{1}\{#1\}}
\newcommand{\xb}{\bar{x}}
\newcommand{\hQ}{\hat{Q}}
\newcommand{\hq}{\hat{q}}
\newcommand{\hT}{\hat{T}}

\newtheorem{theorem}{Theorem}
\newtheorem{lemma}{Lemma}
\newtheorem{corollary}{Corollary}

\newtheorem{remark}{Remark}

\newtheorem{proposition}{Proposition}
\newenvironment{Proof}[1]{\medskip\par\noindent{\bf Proof:\,}\,#1}{{\mbox{\,$\blacksquare$}\par}}
\newcommand{\figref}[1]{\figurename~\ref{#1}}
\IEEEoverridecommandlockouts
\allowdisplaybreaks

\title{Power Adjusting Games Under Block-Dependent Pool Payout Schemes}

\author{Mustafa Doger \qquad Sennur Ulukus\\
\normalsize Department of Electrical and Computer Engineering\\
\normalsize University of Maryland, College Park, MD 20742\\
\normalsize  \emph{doger@umd.edu} \qquad \emph{ulukus@umd.edu}}

\date{}

\begin{document}

\maketitle

\begin{abstract}
    Advanced power adjusting block withholding attacks such as PAW and T-PAW have been analyzed so far under a global-reset version of the proportional scheme (GRP), which is rarely used in practice. In this paper, we analyze T-PAW, which contains FAW and PAW as special cases, under the block-dependent payout schemes deployed in practice: Pay-Per-Last-N-Shares (PPLNS), round-based Proportional (PROP) and exponential-decay (EXP). We derive the revenue ratios analytically, by renewal arguments for PPLNS and through the stationary carry process of an open round for PROP and EXP. We further introduce Power Adjusting Hopping (PAH), where an adversary of any size adjusts its power allocation towards the pool according to the blockwork age of the round without withholding any block, and derive its revenue ratios under PROP and EXP. We show that GRP misestimates the T-PAW gain in a scheme-dependent direction: it overstates the gain under PROP up to an order of magnitude and understates it under EXP for the original slush decay rate. PROP is highly vulnerable to PAH, while EXP faces a trade-off: a slow decay invites PAH and a fast decay invites T-PAW. PPLNS is the least vulnerable scheme: it is immune to PAH and it mitigates, but does not eliminate, T-PAW. All analytical results are verified by Monte Carlo simulations.
\end{abstract}

\section{Introduction}
Miners in Nakamoto style blockchains such as Bitcoin try to solve cryptographic puzzles to append blocks to the longest chain and earn rewards~\cite{btc-whitepaper}. For a single miner, as the global hashrate grows, it becomes statistically harder to solve Proof-of-Work (PoW) puzzles hence miners cooperate under pools and share the reward of the blocks they create. Pools distribute rewards of blocks according to their choice of payout schemes. Block-dependent schemes Proportional (PROP), Pay-Per-Last-N-Shares (PPLNS) and score-based such as Exponential Decay (EXP) pay the partial works (pPoW) when pool finds a block (fPoW), whereas share-based schemes like Pay-Per-Share (PPS) immediately pay each pPoW~\cite{rosenfeld2011analysisbitcoinpooledmining}.

While honest miners always try to extend the longest chain, an adversary may consider alternative strategies such as selfish mining where a miner temporarily withholds the blocks it mines to increase its relative share of rewards~\cite{selfish-mining}. Similarly, in pools with block-dependent schemes, a miner can withhold fPoWs from the manager and strategically split its hashpower between the pool and solo mining to increase its relative rewards, called \textit{Block Withholding} (BWH) attack \cite{courtois2014subversiveminerstrategiesblock,power_splitting_pools}. Both selfish mining and BWH strategies have received significant interest for optimization purposes from different perspectives~\cite{optimal-selfish,stubborn-mining,intermittent_mining,optimal-bwh,miners_dilemma,fork_after_witholding_attack,power_adjusting,doger2026temporarypoweradjustingwithholding,return-after-withholding,profit_lag,uncle_block_attack,Wang2023EFAW,if_you_cant_beat_pay_them,bm-paw_attack,Zhu2022_revisit_faw,doger2026alloutattackoptimalblock,block_withholding_delay}. \textit{Fork After Withholding} (FAW)~\cite{fork_after_witholding_attack}, \textit{Power Adjusting Withholding} (PAW)~\cite{power_adjusting}, and \textit{Temporary PAW} (T-PAW)~\cite{doger2026temporarypoweradjustingwithholding} successively added flexibility to BWH for this purpose. Recently,~\cite{doger2026alloutattackoptimalblock} showed that share-based schemes such as PPS are vulnerable to All-out Attack (AoA, a special type of BWH), contrary to earlier belief \cite{Saide_Zhu_2018_Mathematical_Foundations_of_Computing}.

Block-dependent round-based schemes also invite another attack family called \emph{pool hopping}: an adversary leaves the pool for solo mining or another pool as the shares accumulate. Rosenfeld~\cite{rosenfeld2011analysisbitcoinpooledmining} and Raulo~\cite{raulo2011optimal} show that a miner with negligible hashpower can increase its profit by $28\%$ under PROP by doing so, and argue qualitatively that score-based schemes such as EXP reduce, but do not remove it. Rosenfeld also shows that PPLNS is hopping-proof as long as the difficulty and the block reward are constant. On the empirical side,~\cite{belotti2018poolhopping,cortesi2022new} propose methods to detect hoppers, find that hoppers earn more than static miners, including under score-based schemes, and~\cite{tovanich2021empirical} shows that payout schemes and pool fees drive the migration of miners across pools. \cite{schrijvers2016incentive,zolotavkin2017incentive,lazos2021rpplns,chang2020sharewithholdingattackblockchain,chen_pplns_delay} study the incentive to delay the report of shares or blocks within a pool.

These works leave three gaps. First, the analyses of advanced withholding attacks~\cite{power_adjusting,return-after-withholding,doger2026temporarypoweradjustingwithholding} pay a released fPoW according to the shares submitted since the last block of the whole network, i.e., a global-reset version of PROP (GRP), which is not used in practice. Since the gain of these attacks stems from changing the power allocation within a round, it is not clear whether the conclusions carry over to the round-based PROP, PPLNS and score-based schemes that pools deploy. Second, the pool hopping analyses~\cite{rosenfeld2011analysisbitcoinpooledmining,raulo2011optimal} assume a miner with negligible hashpower that either mines fully for a pool or solo and the vulnerability of EXP to hopping has not been analyzed in closed-form, although hopping under score-based schemes is observed in practice~\cite{belotti2018poolhopping}. Third, withholding and hopping have been studied separately, thus it is not known which attack is more effective under each scheme, or how the block-dependent schemes compare with PPS, which is shown to be vulnerable to BWH~\cite{doger2026alloutattackoptimalblock}. We address these gaps with the following contributions (also see Table~\ref{tab::gap} and~\ref{tab::summary}):
\begin{itemize}
    \item We analyze T-PAW, which contains FAW and PAW as special cases, under PPLNS, round-based PROP and EXP. We derive the reward fractions by renewal arguments for PPLNS, and through the stationary carry process of the open round for PROP and EXP, and we show that GRP follows as a special case of PROP with zero carry.
    \item We introduce Power Adjusting Hopping (PAH), where an adversary of any size changes its power allocation according to the blockwork age of the round. We derive its revenue ratios in closed-form under PROP and EXP, recover the classical $28\%$ gain as $\alpha\to0$.
    \item We show that GRP misestimates the T-PAW gain: it overstates the gain under PROP up to an order of magnitude and understates it under EXP.
    \item We show that each scheme is vulnerable to a different attack: PROP is dominated by PAH, EXP is vulnerable to both attacks with a trade-off in the decay rate $\lambda_d$, and PPLNS is immune to PAH and yields the smallest T-PAW gain among all schemes, including PPS. We verify all analytical results with Monte Carlo simulations.
\end{itemize}

\begin{table}[t]
    \centering
    \small
    \renewcommand{\arraystretch}{1.25}
    \setlength{\tabcolsep}{3pt}
    \begin{tabularx}{\columnwidth}{|>{\centering\arraybackslash}p{1.3cm}|>{\centering\arraybackslash}X|>{\centering\arraybackslash}X|}
    \hline
    \textbf{Topic} & \textbf{Prior work} & \textbf{This paper}\\
    \hline
    BWH variants & FAW, PAW, T-PAW under GRP~\cite{fork_after_witholding_attack,power_adjusting,doger2026temporarypoweradjustingwithholding} & T-PAW under PPLNS, PROP, EXP; GRP is not a reliable proxy\\
    \hline
    Hopping, PROP & Negligible miner, all-or-nothing~\cite{rosenfeld2011analysisbitcoinpooledmining,raulo2011optimal} & PAH for any $\alpha$ and partial allocations\\
    \hline
    Hopping, EXP & Qualitative~\cite{rosenfeld2011analysisbitcoinpooledmining}, observed in practice~\cite{belotti2018poolhopping} & First closed-form analysis, trade-off in $\lambda_d$\\
    \hline
    Hopping, PPLNS & Hopping-proof~\cite{rosenfeld2011analysisbitcoinpooledmining} & Any $\alpha$ and policy\\
    \hline
    PPS & BWH profitable~\cite{doger2026alloutattackoptimalblock} & Compared with all block-dependent schemes\\
    \hline
    \end{tabularx}
    \caption{Prior work and the gaps addressed in this paper.}
    \label{tab::gap}
\end{table}

\begin{table}[t]
    \centering
    \footnotesize
    \renewcommand{\arraystretch}{1.25}
    \setlength{\tabcolsep}{2.5pt}
    \begin{tabularx}{\columnwidth}{|c|>{\centering\arraybackslash}X|>{\centering\arraybackslash}X|>{\centering\arraybackslash}X|}
    \hline
    \textbf{Scheme} & \textbf{Withholding} & \textbf{Hopping (PAH)} & \textbf{Overall}\\
    \hline
    PPS & High (AoA~\cite{doger2026alloutattackoptimalblock}) & Immune & Vulnerable, loss on operator\\
    \hline
    PPLNS & Low (T-PAW) & Immune & Most robust\\
    \hline
    PROP & Low (T-PAW) & Very high & Vulnerable to PAH\\
    \hline
    EXP & Moderate (T-PAW), grows with $\lambda_d$ & Moderate, decreases with $\lambda_d$ & No $\lambda_d$ protects\newline against both\\
    \hline
    GRP & Overstates PROP, understates EXP & -- & Not reliable/used in practice\\
    \hline
    \end{tabularx}
    \caption{Summary of the vulnerability of pool payout schemes to withholding and hopping attacks.}
    \label{tab::summary}
\end{table}

\section{System Model}\label{sec::model}

\begin{table}[t]    
    \begin{center}
    \begin{tabular}{||c | c||} 
    \hline
    Parameter & Description/Value \\ [0.5ex] 
    \hline\hline
    $\lambda_1$ & total block generation rate \\
    \hline 
    $\alpha$ & adversarial hash fraction  \\ 
    \hline
    $\beta$ & target pool hash fraction  \\ \hline
    $\gamma$ & fraction of outside honest miners \\&on the adversarial fork\\
    \hline
    $\mathds{1}_{P_R}$ & 1 if the pool manager is rational\\
    \hline
    $T$ & withholding deadline (T-PAW)\\
    \hline
    $t^*$ & hopping threshold (PAH)\\
    \hline
    $W$ & PPLNS blockwork window\\
    \hline
    $\lambda_d$ & EXP decay rate \\
    \hline
    $p_i$ ($i=1,2,3,y,o$) & fraction of $\alpha$ at pool mining\\
    \hline
    $q_i,\ a_i,\ a'_i$&$1-p_i,\ \beta+\alpha p_i,\ \frac{\alpha p_i}{a_i}$\\
    \hline
    $\xi$&$1-\alpha-\beta$\\
    \hline
    $\pi_A,\ \pi_P,\ \pi_R$ & $\frac{\alpha q_2}{1-\alpha p_2},\ \frac{\beta}{1-\alpha p_2},\ \frac{\xi}{1-\alpha p_2}$  \\
    \hline
    $c$&$\gamma\xi+\alpha q_3+a_3\mathds{1}_{P_R}$\\
    \hline
    $\lambda_2,\ \mu_2$ & $(1-\alpha p_2)\lambda_1,\ \frac{\lambda_2}{\lambda_1a_2}$  \\
    \hline
    $\kappa_o,\ \kappa_1$ & $\frac{\lambda_1a_o}{\lambda_d},\ \frac{\lambda_1}{\lambda_d}$\\
    \hline
    $\theta,\ \hT,\ \chi_T$&$e^{-\lambda_2T},\ \lambda_1a_2T,\ e^{-\lambda_dT}$\\
    \hline
    $y^*,\ e^*,\ \chi^*$ & $\lambda_1a_yt^*,\ e^{-y^*},\ e^{-\lambda_dt^*} $\\
    \hline
    $E_1(t)$ & $\int_t^{\infty}\frac{e^{-x}}{x}dx$\\
    \hline
    \end{tabular}
    \end{center}
    \caption{ Frequently used notations.}
    \label{table::freq-table}\vspace{-.5cm}
\end{table}
We study a Nakamoto-style blockchain secured by PoW, where miners extend the longest chain. The hashpower is spread over many miners, each holding a negligible share, and the whole network finds blocks as a Poisson process of rate $\lambda_1$, e.g., $\lambda_1=1/600$ per second in Bitcoin. Hence, any party that controls a fraction $x$ of the hashpower finds blocks as an independent Poisson process of rate $x\lambda_1$.  Table~\ref{table::freq-table} summarizes the notation used throughout the paper.

\emph{Participants:} A coalition of colluding miners that controls a fraction $\alpha$ of the hashpower acts as a single entity, which we call the adversary. A single mining pool, whose members are all honest and which controls a fraction $\beta$ of the hashpower, is the target of the adversary, and the remaining fraction $\xi=1-\alpha-\beta$ belongs to honest miners outside the target pool. Honest miners always build on the tip of the longest chain they know and publish every block they find without delay. We assume $\alpha+\beta<\frac{1}{2}$. The adversary may divide its hashpower between the target pool and mining on its own; we refer to the latter as \emph{solo} mining, during which the adversary follows the protocol and publishes each block it finds immediately.

\emph{Pool mining:} The manager of the target pool distributes work to its members, who report two kinds of \emph{shares}. fPoWs are valid blocks, while pPoWs satisfy a lower difficulty and serve as evidence of the work done. If $d_p$ and $d_f$ denote the pPoW and fPoW difficulties, each share is also a fPoW with probability $d_p/d_f$, independently of other shares. We quantify hashing in units of \emph{blockwork}, the amount of hashing that yields one fPoW in expectation. A party with hash fraction $x$ performs the blockwork $x\lambda_1t$ during a time $t$, the number of fPoWs it finds is Poisson with this mean, and each share carries the blockwork $d_p/d_f$. The manager divides the block reward among the members according to a payout scheme.

\emph{Forks:} In line with the literature on incentive attacks~\cite{selfish-mining,fork_after_witholding_attack}, we assume that the adversary observes every block as soon as it is found, and we neglect the accidental forks among honest blocks caused by propagation delays. Forks therefore arise only from the adversary's actions. When the adversary publishes a withheld fPoW in response to a block of an honest miner outside the pool, a $\gamma$ fraction of the outside honest miners continue on the adversarial block and the remaining $1-\gamma$ fraction on the honest one. The members of the target pool follow the choice of their manager. We write $\mathds{1}_{P_R}=1$ if the manager is rational, i.e., it builds on the fPoW released into its own pool and $\mathds{1}_{P_R}=0$ otherwise.

\emph{Metrics:} A block is \emph{canonical} if it remains in the longest chain in the long run. The \emph{block redundancy ratio} $\delta$ is the expected number of all blocks found, including orphaned and discarded ones, per canonical block. As in Bitcoin, a difficulty adjustment algorithm keeps the rate of canonical blocks constant in the long run, hence the long-run income of an entity is proportional to its share of the canonical block rewards. Accordingly, we denote by $\rho_A$, $\rho_P$ and $\rho_R$ the \emph{revenue ratios}, i.e., the long-run fractions of the canonical block rewards received by the adversary, the honest members of the target pool and the honest miners outside the pool, respectively, with $\rho_A+\rho_P+\rho_R=1$. Under honest mining, $(\rho_A,\rho_P,\rho_R)=(\alpha,\beta,\xi)$, and we measure the gain of an attack by the \emph{relative extra reward} of the adversary $\mathrm{RER}_A=\frac{\rho_A}{\alpha}-1$.

\subsection{Block-Dependent Payout Schemes}\label{sec::block_dep_pay_scheme}
A block-dependent payout scheme is one where the miners are paid only when the pool actually discovers a valid block. Although shares do not get direct payments, the final payment a miner receives when a block is discovered is a function of the share contributions. Since miners only get payments when the pool finds blocks, the cashflow of miners is less stable compared to share-based schemes such as PPS, where each valid share submitted to pool operator gets an immediate payment from the pool operators that is valued at a fraction of the coinbase reward. Hence, in block-dependent schemes the risk associated with the variance of block mining process is borne by the pool miners collectively. These schemes are attractive because pools need smaller reserves, operator risk is much lower and as a result miners face lower fees. 

In this paper, we focus on mainstream block-dependent mining pool payout schemes: PROP, PPLNS and EXP, where miners are paid only when the pool actually discovers a valid block. These three schemes cover the main design ideas, and together they span the space of block-dependent schemes well. Share-based schemes are not affected by power adjusting within the pool, since the payment of a share does not depend on the rounds; however, they are vulnerable to BWH after the difficulty adjustment~\cite{doger2026alloutattackoptimalblock}, and we compare them with the block-dependent schemes in Section~\ref{sec::numerical}.

\subsubsection{Pay-Per-Last-N-Shares ($W$-PPLNS)}\leavevmode\\
PPLNS is a sliding-window payout scheme: when the pool finds a block, the block reward $R_b$ is split proportionally among the last $N$ shares submitted to the pool, i.e., a miner who contributed $n$ of these $N$ shares receives $R_b\frac{n}{N}$. We parametrize the window by its blockwork $W\triangleq Nd_p/d_f$ and refer to the scheme as $W$-PPLNS since $N$ and $W$ are in one-to-one correspondence for given difficulties. In real-world PPLNS implementations, $N$ typically covers a small number of blockworks~\cite{belotti2018poolhopping}, e.g., $W=2$ \cite{p2pool_stats_20261008,chernykh2023p2poolv3}.  

\subsubsection{Proportional (PROP)}\leavevmode\\
PROP is a simple round-based payout scheme where a block discovery triggers reward split proportionally among all shares submitted during that round. In this context, a round refers to the time interval between two consecutive pool blocks that become canonical. In PROP, a miner who contributes $n_s$ shares out of the $n_S$ total shares the pool receives during the round, receives $R_b\frac{n_s}{n_S}$ rewards at the end of the round.

\subsubsection{Score-based: Exponential-decay (EXP)}\leavevmode\\ 
Score-based schemes are mining-pool reward systems where each share gets a time-dependent weight that gradually decreases as it gets older. In this paper, we specifically focus on historically deployed slush-style round-based modeling with exponential-decay, where the shares within a round are weighted with an exponential function $w(t)=e^{-\lambda_d t}$, where $\lambda_d$ is the decay rate, which we express relative to the total block generation rate as $\frac{\lambda_1}{\lambda_d}=\kappa_1$ (e.g., $\kappa_1=1/2$ in the original slush method~\cite{rosenfeld2011analysisbitcoinpooledmining,bitcoinwiki_pools_expdecay}). More specifically, when a round ends, each share $i$ that is mined $t_i$ time ago within that round by a miner $j$ gets a weight $w(t_i)=e^{-\lambda_dt_i}$ and miner $j$ receives $\frac{\sum_{i\in j}w(t_i)}{\sum_{k}w(t_k)}$ reward at the end of the round . 

\subsubsection{Global-reset-PROP (GRP)}\leavevmode\\
Recent analyses of advanced block withholding attacks~\cite{power_adjusting,return-after-withholding,doger2026temporarypoweradjustingwithholding} implicitly adopt an unconventional version of PROP scheme that is rarely, if ever, used in practice, which we call GRP. Under GRP, rounds start and end whenever any block becomes known to the pool manager, whether it is mined by the pool or by an entity outside the pool. Since only the shares submitted since the last block of the network are paid, pool miners bear considerably more risk than under PROP. On the other hand, GRP is more appealing from an academic perspective, as it is considerably cleaner to analyze. As we show later, this simplification is not innocuous: GRP gives more weight to recent shares and therefore overstates the gain of power adjusting attacks compared to the round-based PROP.

\section{Power Adjusting Withholding Attacks under Block-Dependent Payout Schemes}\label{sec::withholding}
In this section, we review the classical BWH attack and its advanced variants, FAW, PAW, and T-PAW. We present the relevant formulas associated with these attacks, including revenue ratios and block redundancy ratios, under the implicit assumption of a block-dependent payout scheme. The revenue ratios depend on the payout scheme only through the adversarial reward fractions of the pool blocks. Then, we provide novel and exact formulas of the reward fractions for all four block-dependent schemes listed in Section~\ref{sec::block_dep_pay_scheme}.

\begin{figure*}[t]
    \centering
    \includegraphics[width=1\textwidth]{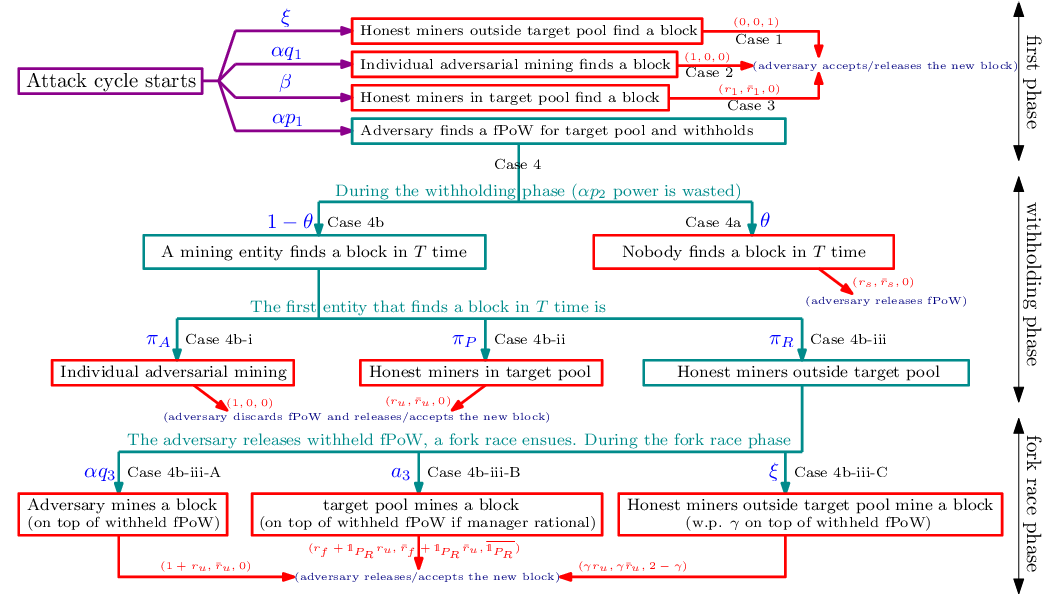}
    \caption{Evolution of a T-PAW attack cycle.}
    \label{fig::cycle_events}
\end{figure*}

\subsection{Classical Block Withholding (BWH)}
Under this attack, the adversary allocates $p$ fraction of its hashpower to the target pool and mines solo with the remaining power. In the target pool, the adversary submits every pPoW it finds to the pool manager but withholds every fPoW it finds. It is straightforward to obtain (with $q=1-p$) the following
\begin{align}
    \rho_{A}&=(\alpha q+\beta r)\delta^{\rm BWH},\ 
    \rho_{P}=\beta \Bar{r}\delta^{\rm BWH},\
    \rho_{R}=\xi\delta^{\rm BWH},\\
    \delta^{\rm BWH}&=\frac{1}{1-\alpha p}.
\end{align}
Here, $r$ is the expected adversarial reward fraction, and $\Bar{r}=1-r$ is the fraction that goes to the honest pool members. Since the adversary never changes its power allocation $p$ under BWH, the adversarial fraction of the pool blockwork is the same at all times, thus for all four block-dependent schemes listed in Section~\ref{sec::block_dep_pay_scheme},
\begin{align}
    r=\frac{\alpha p}{\beta+\alpha p},\label{eq::r_BWH}
\end{align}
regardless of which past shares the scheme pays. Subsequently, the adversary can optimize an objective function, such as its revenue ratio, by adjusting $p$.

\subsection{Advanced Variants}
Here, we focus on three natural extensions of BWH, namely FAW~\cite{fork_after_witholding_attack}, PAW~\cite{power_adjusting}, and T-PAW~\cite{doger2026temporarypoweradjustingwithholding}. On a high level, the extensions enable us to state that FAW $\subset$ PAW $\subset$ T-PAW in terms of the attack space, hence we directly present the T-PAW scheme. In their analysis, both \cite{fork_after_witholding_attack,power_adjusting} use a simplified model for fork races whereas \cite{doger2026temporarypoweradjustingwithholding} uses a more realistic version by considering the fork races explicitly using $\gamma$, which we adopt in this paper. However, both \cite{power_adjusting,doger2026temporarypoweradjustingwithholding} use the GRP payout scheme, as it is considerably easier to analyze, but it overstates the gain of their power adjusting attacks. T-PAW also contains the lie-in-wait attack of~\cite{rosenfeld2011analysisbitcoinpooledmining}, in which the adversary temporarily withholds a fPoW while increasing its power allocation to the pool~\cite{doger2026temporarypoweradjustingwithholding}. Notably, lie-in-wait was proposed for PPLNS pools and analyzed only approximately, i.e., for a small withholding duration and without fork races. Hence, in this paper, we introduce a general framework to analyze the T-PAW attack, which spans all these attack spaces, for all four block-dependent schemes listed in Section~\ref{sec::block_dep_pay_scheme}.

First, we explain how a T-PAW attack cycle works which contains FAW and PAW attacks as special cases. To avoid repetition, when we explain the attack cycle in the next paragraph, we advise the reader to follow \figref{fig::cycle_events}, which contains all the relevant cases and phases, as well as the corresponding probabilities and rewards. Note that, in \figref{fig::cycle_events} an attack cycle starts with the purple box and ends with any of the events listed in the red box. The arrows represent the transition to subcases and blue symbols near the arrows show the probabilities. Each red box ends the cycle and points to the rewards each entity gets from the cycle as a 3-tuple $(\cdot ,\cdot ,\cdot )$ where the fields are the rewards of the adversary, the honest target pool miners and the remaining honest miners respectively.

Let us assume everyone agrees on the last block mined in the network and a new attack cycle starts, where the adversary mines solo with $q_1$ fraction of its hashpower and contributes to the pool with the remaining $p_1$ fraction. This phase is called \emph{first phase} and the first miner to mine a block is
\begin{itemize}[leftmargin=0cm]
    \item[] Case~1: Honest miners outside the target pool. The adversary accepts the block and the cycle ends.
    \item[] Case~2: Solo adversary. The adversary releases the block and the cycle ends.
    \item[] Case~3: Honest miners of the target pool. The adversary accepts the block and the cycle ends.
    \item[] Case~4: The adversarial power in the target pool. The adversary withholds the fPoW and readjusts its power contributing to the target pool as $p_1\rightarrow p_2$. This phase is called \emph{withholding phase} where the adversary continues mining; however, if it encounters another fPoW with its pool power $p_2$, it discards the old fPoW and keeps the new one. If 
    \begin{itemize}[leftmargin=0.2cm]
        \item[] Case~4a: No entity other than adversarial pool mining power finds a block for $T$ time duration, the adversary releases the withheld fPoW and attack cycle ends.
        \item[] Case~4b: The first entity that finds a block in $T$ time is:
        \begin{itemize}[leftmargin=0.2cm]
            \item[] Case~4b-i: Solo adversary. The adversary discards its withheld fPoW and releases the solo block and the cycle ends.
            \item[] Case~4b-ii: Honest miners of the target pool. The adversary discards its withheld fPoW and accepts the new pool block and the cycle ends.
            \item[] Case~4b-iii: Honest miners outside the target pool. The adversary releases its withheld fPoW and a fork race ensues. During the \emph{fork race phase}, the adversary readjusts its power contributing to the target pool as $p_2\rightarrow p_3$ and tries to mine on top of the fPoW it recently released. The first entity that finds a block during the \emph{fork race phase} is
            \begin{itemize}[leftmargin=0.2cm]
                \item[] Case~4b-iii-A: Solo adversary. The adversary releases the block and the cycle ends.
                \item[] Case~4b-iii-B: Honest miners of the target pool or the adversarial pool mining. Note that if the target pool manager is rational, this pool block is mined on top of the recently released fPoW by the adversary. In either case, the adversary accepts/releases the block and the cycle ends. 
                \item[] Case~4b-iii-C: Honest miners outside the target pool. Note that $\gamma$ fraction of these miners prefer the recently released fPoW by the adversary. In either case, the adversary accepts/releases the block and the cycle ends. 
            \end{itemize}
        \end{itemize}
    \end{itemize}
\end{itemize}

At the end of the attack cycle, the adversary reverts its power allocation towards the target pool as $p_1$ and a new attack cycle starts. Note that in comparison to \cite{doger2026temporarypoweradjustingwithholding}, we introduced another flexibility to the \emph{fork race phase} of T-PAW for the adversary which can readjust its power allocation towards the target pool as $p_2\rightarrow p_3$. This choice does not necessarily bring major improvements; however, it makes FAW, where the adversary never changes its allocation, an exact special case of T-PAW with $p_1=p_2=p_3$, under which all adversarial reward fractions reduce to $r$. Note that, corresponding expressions for PAW (by taking $T \to \infty$) and FAW (by taking $T \to \infty$ and setting $p_3=p_2 = p_1$) can be obtained.

It is fairly straightforward to verify each of the probabilities listed in \figref{fig::cycle_events}. Regarding the 3-tuple rewards, note that a pool block released at the first phase is distributed between the adversary and the honest pool members as $(r_1,\Bar{r}_1)$. In Case~4a, the reward fractions of the released fPoW are $(r_s,\Bar{r}_s)$. On the other hand, in Case~4b, the honest pool block of Case~4b-ii and the released fPoW of Case~4b-iii, if it ends up being canonical, both have the reward fractions $(r_u,\Bar{r}_u)$. Note that, the pool may mine an extra block in Case~4b-iii-B (during the fork race), which has reward fractions $(r_f,\Bar{r}_f)$. Using this parametrization and $\bar r_x=1-r_x$ with $x\in\{1,s,u,f\}$ it is straightforward to verify the 3-tuple rewards of each case as shown in \figref{fig::cycle_events}. It remains to combine the total rewards of each case with the corresponding probability and then explicitly derive the reward fractions $r_1,r_s,r_u,r_f$.  

During an attack cycle, let $B_A$, $B_P$ and $B_R$ denote the reward obtained by the adversary, honest target pool miners and honest miners outside the target pool, respectively. Let $B_C$ and $B_O$ denote the total numbers of canonical blocks and overall blocks generated during an attack cycle, respectively. Combining the rewards and probabilities of each case in \figref{fig::cycle_events}, which extends~\cite{doger2026temporarypoweradjustingwithholding} with an additional parameter $p_3$, we have
\begin{align}
    \rho_x=&\frac{\mathbb{E}[B_x]}{\mathbb{E}[B_C]},
\end{align}
where $x\in\{A,P,R\}$ with
\begin{align}
    \mathbb{E}&[B_C]=1+\alpha p_1(1-\theta)\pi_R,\\
    \mathbb{E}&[B_A] =  \alpha q_1 + \beta r_1 + \alpha p_1\bigg(\theta r_s +\big(1-\theta \big)\nonumber \\
    &\times\Big( \pi_A + \pi_P r_u+\pi_R \big(c r_u+a_3 r_f +\alpha q_3\big)\Big)\bigg),\\
    \mathbb{E}&[B_P] = \beta \Bar{r}_1 + \alpha p_1\Bigg(\theta \Bar{r}_s+\big(1-\theta \big) \nonumber \\
    &\times\Big( \pi_P \Bar{r}_u + \pi_R\big(c\Bar{r}_u+a_3 \Bar{r}_f\big)\Big)\Bigg),\\
    \mathbb{E}&[B_R] = \xi\bigg(1+\big(1-\theta \big)\frac{\alpha p_1}{1-\alpha p_2}\big(\xi(2-\gamma)+a_3 \overline{\mathds{1}_{P_R}}\big)\bigg),
\end{align}
Here, we note that the reward fractions of a pool block may depend on blockwork submitted in earlier attack cycles, hence $\E[B_A]$ and $\E[B_P]$ are understood as long-run (stationary) expectations per attack cycle, and $r_1,r_s,r_u,r_f$ as the corresponding stationary adversarial reward fractions. Further, since the events that end an attack cycle are races between independent exponential clocks, their probabilities and the number of blocks they create do not depend on how the pool shares rewards. Hence, $\E[B_C]$, $\E[B_R]$, $\E[B_O]$, and therefore $\rho_R$ and $\delta$, are the same for every block-dependent payout scheme and do not depend on the adversarial reward fractions $r_1,r_s,r_u,r_f$. On the other hand, $\E[B_A]$, $\E[B_P]$ keep their form but the adversarial reward fractions $r_1,r_s,r_u,r_f$ depend on the scheme, i.e., the payout scheme only moves reward between the adversary and the honest pool members. The exact values of these adversarial reward fractions are provided later for each scheme.  It is straightforward to verify that $\rho_R\geq\xi$ and $\rho_A+\rho_P+\rho_R=1$. Further, the block redundancy ratio is
\begin{align}
    \delta^{\rm T-PAW}&=\frac{\mathbb{E}[B_O]}{\mathbb{E}[B_C]}\label{eq::withhold_cycle_dur_type_b},
\end{align}
where
\begin{align}
    \mathbb{E}[B_O]=&\mathbb{E}[B_C]+\alpha p_1\Bigg(\theta \alpha p_2 \lambda_1 T\nonumber\\&+\big(1-\theta \big)\bigg(1+\alpha p_2 \lambda_1 f_{\lambda_2}(T)\bigg)\Bigg),
\end{align}
with $f_{\lambda}(t)=\frac{1}{\lambda}-\frac{ t e^{-\lambda t}}{1-e^{-\lambda t}}$. The goal of the adversary is to optimize an objective function, such as the revenue ratio $\rho_{A}$, with respect to $p_1$, $p_2$, $p_3$ and $T$. In general, objective functions such as $\rho_{A}$ are neither concave nor convex.

\subsection{Reward Fractions in Advanced Variants}
For every scheme, the adversarial reward fraction of a pool block is the adversarial fraction of the shares that the scheme pays for that block, i.e., the shares in the last $W$ blockworks under PPLNS, the shares of the round under PROP, the decay-weighted shares of the round under EXP, and the shares since the last block of the network under GRP. Since the adversary submits shares with different fractions $a_1',a_2',a_3'$ in different phases, these fractions depend on which past phases the scheme pays and with what weight.

First, note that $r_s$ does not appear in the final expression for FAW and PAW since $T \to \infty$ implies $\theta=0$. Further, the analysis does not depend on the particular values of $T$ and $\lambda_1$ but only on the product $\lambda_1 T$. In other words, the adversary can first pick $\lambda_1=1$, find the optimal $T$ and rescale it with the actual block production rate $\lambda_1$. Moreover, the adversarial reward fractions $r_s$ and $r_u$ are affected by the adversarial power adjustments towards the target pool during the withholding phase and deviate from $r_1$ accordingly. This deviation is more pronounced in schemes where recent shares have higher weight, such as EXP or GRP, which makes them more susceptible to PAW/T-PAW as we will see later.
\subsubsection{Reward Fractions in FAW}\leavevmode\\
In FAW, the adversary does not alter its power allocation devoted to pool mining, i.e., $p_i=p$ for $i=1,2,3$ and $T\to\infty$. It is straightforward to show that the adversarial reward fraction for FAW under all four block-dependent payout schemes listed in Section~\ref{sec::block_dep_pay_scheme} remains equal to $r$ given in~\eqref{eq::r_BWH}; that is,
\begin{align}
    r^{FAW}_1 = r^{FAW}_u =r^{FAW}_f = r.
\end{align}

\subsubsection{Reward Fractions for $W$-PPLNS}\leavevmode\\
Let $S$ be the blockwork of the target pool in an attack cycle. An analysis of $W$-PPLNS scheme is provided in Appendix~\ref{app::pplns_scheme} using renewal equations and the following result is obtained.

\begin{theorem}\label{thm:closed_pplns}
    With $W\gg\E[S]$, up to vanishing terms,
    \begin{align}
        r_1=&\xb+\frac{1}{W}\Big[(a_1'-\xb)\,a_1+d\Big],\label{eq:r1}\\
        r_s=&\bar{x}+\frac{1}{W}\Big[(a_1'-\bar{x})\,a_1+(a_2'-\bar{x})\,\hT+d\Big],\label{eq:rs}\\
        r_u=&\xb+\frac{1}{W}\Big[(a_1'-\xb)\,a_1+(a_2'-\xb)\,\nu_1+d\Big],\label{eq:ru}\\
        r_f=&\xb+\frac{1}{W}\Big[(a_1'-\xb)\,a_1+(a_2'-\xb)\,\nu_1+(a_3'-\xb)\,a_3+d\Big],\label{eq:rn}
    \end{align}
    with $\xb$ in \eqref{eq:xbar}, $d$ in \eqref{eq:d} and $\nu_1$ in \eqref{eq:nu1}. Here, $\xb$ is the long-run adversarial fraction of the pool blockwork, and the $1/W$ terms correct it for the blockwork of the current attack cycle, which lies at the end of the window and has the adversarial fraction $a_i'$ in phase $i$. The exact expressions are given in Theorem~\ref{thm:exact}.
\end{theorem}

\subsubsection{Reward Fractions for PROP}\leavevmode\\
An analysis of PROP scheme is provided in Appendix~\ref{app::prop_scheme} using a carry process that models how the blockwork of attack cycles that produce no canonical pool block is carried over to the next cycles. With the help of the Laplace transform of the carry process, the following result is obtained.

\begin{theorem}\label{thm::round_prop}
    The expected adversarial reward fractions under round-based PROP are, for $x\in\{1,s,u,m\}$,
    \begin{align}
        r_x&=a_1'+(a_2'-a_1')\,\omega_2^{x}+(a_3'-a_1')\,\omega_3^{x},\label{eq::r_x}\\
        r_f&=\mathds{1}_{P_R}\,a_3'+\overline{\mathds{1}_{P_R}}\,r_m.\label{eq::r_f}
    \end{align}  
    Here, $r_m$ is the adversarial reward fraction of the pool block of the fork race when the pool manager is not rational, $\omega_i^x\in[0,1]$ is the expected fraction of the total blockwork of the round submitted in mode $i$ (first phase, withholding phase and fork race phase for $i=1,2,3$), both derived in Appendix~\ref{app::prop_scheme}.
\end{theorem}

\subsubsection{Reward Fractions for EXP }\leavevmode\\
Under EXP, the reward of a round depends on the score $\mathbf{Z}=(Z_A,Z)$ of its shares, which is carried over from earlier attack cycles as under PROP, but discounted by $e^{-\lambda_dt}$ over time. An analysis of this scheme is provided in Appendix~\ref{app::exp_scheme}, and the following result is obtained.
\begin{theorem}[Informal]\label{thm::exp_informal}
    The expected adversarial reward fractions under round-based EXP are
    \begin{align}
        r_x&=\int_0^\infty\mathcal{L}_A^{x}(s)\,ds,\qquad x\in\{1,s,u,m\},\\
        r_f&=\mathds{1}_{P_R}\,a_3'+\overline{\mathds{1}_{P_R}}\,r_m,
    \end{align}
    where $\mathcal{L}_A^{x}(s)=\E[Z_Ae^{-sZ}]$ for the score of the corresponding paid round, given in Theorem~\ref{thm::exp} in terms of the unique bounded solution of the linear fixed point equation \eqref{eq::exp_fixed}.
\end{theorem}

Unlike PROP, the transforms are not explicit, but every operator in \eqref{eq::exp_fixed} is a one-dimensional integral, thus \eqref{eq::exp_fixed} is solved on a grid and $r_x$ follows by one-dimensional quadrature; see Theorem~\ref{thm::exp} for the formal statement.

\subsubsection{Reward Fractions for GRP}\leavevmode\\
This scheme is rigorously studied in~\cite{doger2026temporarypoweradjustingwithholding} with $p_3=0$. Since every block of the network starts a new round under GRP, no blockwork is carried over from earlier attack cycles. Hence, GRP is a special case of round-based PROP with zero carry, and Theorem~\ref{thm::round_prop} gives (see Remark~\ref{rem::GRP} in Appendix~\ref{app::prop_scheme}),
\begin{align}
    r_1=&\ a_1',\label{eq::GRP_r1}\\
    r_s=&\ a_1'+(a_2'-a_1')\,\frac{\hT}{a_1}e^{\hT/a_1}E_1\Big(\frac{\hT}{a_1}\Big),\label{eq::GRP_rs}\\
    r_u=&\ a_1'+\frac{a_2'-a_1'}{1-\theta}\int_0^{\hT}\mu_2e^{-\mu_2y}\,\frac{y}{a_1}e^{y/a_1}E_1\Big(\frac{y}{a_1}\Big)dy,\label{eq::GRP_ru}\\
    r_f=&\ a_3'.\label{eq::GRP_rf}
\end{align}

\section{Power Adjusting Hopping Attacks under Block-Dependent Payout Schemes}\label{sec::hop}
In this section, we consider PAH attack that the adversary can use to increase its rewards. A primitive version of this attack for an adversary with negligible hashpower with optimal time to hop under PROP were derived in~\cite{rosenfeld2011analysisbitcoinpooledmining,raulo2011optimal}. Here, we extend the analysis of PAH to PROP, EXP and to any adversarial power size, where we allow partial power allocations. Under PAH, the adversary never withholds a fPoW, i.e., it submits all its pPoWs and fPoWs to the pool immediately, and only adjusts the fraction of its hashpower that it devotes to the target pool. Since every pool block is released immediately, every block is canonical and the round boundaries of the pool are simply the instants at which the pool finds a block. The adversary exploits the fact that, under round-based PROP and EXP, the value of a share depends on the cumulative blockwork of the open round. 

Let $t$ denote the time passed since the last pool block, which is public. Under PAH, the adversary mines in the pool with the fraction
\begin{align}
    p(t)=p_y\,\ind{t<t^*}+p_o\,\ind{t\geq t^*}\label{eq::hop_policy}
\end{align}
of $\alpha$ and solo with the rest, where $p_y,p_o\in[0,1]$ and $t^*\geq0$ are the parameters of the attack; the subscripts $y$ and $o$ refer to young and old rounds. Similarly, let $a(t)$, $a'(t)$, $q(t)$ denote the quantities $a_j,a'_j,q_j$ when the adversary mines with the power $p(t)=p_j$ with $j\in\{y,o\}$ at time $t$. Note that with $p_y=p_o=p$, the strategy reduces to the honest strategy that keeps the fraction $p$ of its hashpower in the pool. Further, let
\begin{align}
    y^*&=\lambda_1a_yt^*,& e^*&=e^{-y^*},\label{eq::hop_notation1}\\
    \chi^*&=e^{-\lambda_dt^*},& \kappa_o&=\frac{\lambda_1a_o}{\lambda_d}.\label{eq::hop_notation2}
\end{align}
We refer to the cumulative pool blockwork of the open round as its \emph{blockwork age}, i.e., the blockwork age at time $t$ is
\begin{align}
    y(t)=\int_0^t\lambda_1a(t')\,dt',\label{eq::hop_age}
\end{align}
and $y(t^*)=y^*$, i.e., the policy \eqref{eq::hop_policy} is equivalently a threshold $y^*$ on the blockwork age of the round. Let $\tau$ be the time when the pool block is found and the round ends, and $Y_\tau\triangleq y(\tau)$ the blockwork age of the round at $\tau$. Since every hash is a valid block with the same probability, the chance that the pool finds a block depends only on how much blockwork it has done and not on how fast it was done. Hence, the power allocation of the adversary changes how fast the blockwork age of a round grows, but not the distribution of $Y_\tau$, which is $\mathrm{Exp}(1)$ for every policy (see Lemma~\ref{lem::hop_round} in Appendix~\ref{app::hop}).

\begin{theorem}\label{thm::hop}
    Under round-based PROP and EXP, the block redundancy ratio of PAH is $\delta^{\rm PAH}=1$ and
    \begin{align}
        \rho_A&=\frac{n_S+r_{h}}{n_S+n_R+1},\label{eq::hop_rho_A}\\ 
        \rho_P&=\frac{\Bar{r}_{h}}{n_S+n_R+1},\label{eq::hop_rho_P}\\
        \rho_R&=\frac{n_R}{n_S+n_R+1},\label{eq::hop_rho_R}
    \end{align}
    where $(r_h,\Bar{r}_h)$, are the reward fractions of the adversary and the honest target pool miners with $\Bar{r}_h=1-r_h$, and
    \begin{align}
        n_R&=\frac{\xi}{a_y}(1-e^*)+\frac{\xi}{a_o}e^*,\label{eq::hop_nr}\\
        n_S&=\frac{\alpha q_y}{a_y}(1-e^*)+\frac{\alpha q_o}{a_o}e^*\label{eq::hop_ns}
    \end{align}
    are the expected numbers of blocks mined by the honest miners outside the pool and the solo adversary per round, with
    \begin{align}
        r_h^{\rm PROP}=&\ a_y'(1-e^*)+a_o'e^*+(a_y'-a_o')\,y^*E_1(y^*),\label{eq::hop_r_prop}\\
        r_h^{\rm EXP}=&\ a_y'(1-e^*)+e^*\kappa_o\int_0^1v^{\kappa_o-1}\nonumber\\
        &\times\frac{\alpha p_o+\big(\alpha p_y(1-\chi^*)-\alpha p_o\big)v}{a_o+\big(a_y(1-\chi^*)-a_o\big)v}\,dv.\label{eq::hop_r_exp}
    \end{align}
\end{theorem}

The proof is provided in Appendix~\ref{app::hop}. For $p_y=p_o=p$, both adversarial reward fractions reduce to $r$ in \eqref{eq::r_BWH}, and Theorem~\ref{thm::hop} gives the revenue ratios of honest mining, $\rho_A=\alpha$, $\rho_P=\beta$ and $\rho_R=\xi$. As $\lambda_d\to0$, EXP weighs all the shares of a round equally and $r_h^{\rm EXP}\to r_h^{\rm PROP}$. The goal of the adversary is to optimize an objective function, such as the revenue ratio $\rho_{A}$, with respect to $p_y$, $p_o$ and $t^*$.

\begin{remark}\label{rem::hop_value}
    Consider a pool share of blockwork $\epsilon\to0$ submitted at blockwork age $y$ under PROP by a negligible miner. In the pool, the share receives the fraction $\epsilon/Y_\tau$ of the reward of the pool block, and given $Y_\tau>y$, the remaining blockwork age $Y_\tau-y$ is $\mathrm{Exp}(1)$ by Lemma~\ref{lem::hop_round}(iii) and the memoryless property. Hence, the expected reward of the share is
    \begin{align}
        r(\epsilon)=\E\bigg[\frac{\epsilon}{Y_\tau}\,\bigg|\,Y_\tau>y\bigg]=\int_0^\infty\frac{\epsilon\, e^{-u}}{y+u}\,du=\epsilon\, e^yE_1(y).\label{eq::hop_share_value}
    \end{align}
    If mined solo, the same blockwork yields $\epsilon$ blocks in expectation, hence the value of the share relative to solo mining is $e^yE_1(y)$ which is exactly what Rosenfeld finds in~\cite[App.~B]{rosenfeld2011analysisbitcoinpooledmining}. It decreases from $\infty$ to $0$, thus shares in young rounds are worth more than solo mining and shares in old rounds less, which is what PAH exploits. EXP reduces, but does not remove, this dependence of the share value on the blockwork age.
\end{remark}

In Appendix~\ref{app::hop}, we revisit the PAH-proofness of $W$-PPLNS \cite{rosenfeld2011analysisbitcoinpooledmining}. Further, we consider a special case with $p_y=1$ and $p_o=0$ and optimize PAH only with respect to $y^*$ for PROP and EXP. For PROP, by taking $\alpha\to 0$, we also recover the classical pool hopping result of~\cite{raulo2011optimal,rosenfeld2011analysisbitcoinpooledmining} for a negligible miner.

\begin{table}[!tp]
    \centering
    \resizebox{\columnwidth}{!}{%
    \begin{tabular}{cc|ccc|ccc|ccc}
    \toprule
     & & \multicolumn{3}{c|}{T-PAW, $\gamma=0$, $\lambda_d=$} & \multicolumn{3}{c|}{T-PAW, $\gamma=0.5$, $\lambda_d=$} & \multicolumn{3}{c}{PAH} \\
    $\alpha$ & $\beta$ & $\lambda_1$ & $2\lambda_1$ & $4\lambda_1$ & $\lambda_1$ & $2\lambda_1$ & $4\lambda_1$ & $\lambda_1$ & $2\lambda_1$ & $4\lambda_1$\\
    \midrule
    0.001 & 0.05 & 0.056 & 0.108 & 0.180 & 0.102 & 0.169 & 0.247 & 3.990 & 2.184 & 1.153\\
    0.001 & 0.1 & 0.035 & 0.061 & 0.098 & 0.060 & 0.092 & 0.130 & 6.921 & 4.010 & 2.195\\
    0.01 & 0.05 & 0.482 & 0.890 & 1.447 & 0.853 & 1.366 & 1.959 & 3.667 & 2.009 & 1.061\\
    0.01 & 0.1 & 0.324 & 0.558 & 0.876 & 0.554 & 0.836 & 1.165 & 6.617 & 3.837 & 2.102\\
    0.05 & 0.05 & 1.438 & 2.382 & 3.579 & 2.312 & 3.425 & 4.666 & 2.768 & 1.523 & 0.805\\
    0.05 & 0.1 & 1.213 & 1.955 & 2.918 & 1.946 & 2.800 & 3.774 & 5.585 & 3.252 & 1.785\\
    \bottomrule
    \end{tabular}}
    \caption{Optimal relative extra reward $\rho_A/\alpha-1$ (\%) under T-PAW and PAH for EXP with different decay rates $\lambda_d$. PAH does not depend on $\gamma$.}
    \label{tab::exp_lambda}
\end{table}

\begin{table}[t]
    \centering
    \setlength{\tabcolsep}{4pt}
    \begin{tabular}{cc|ccc|ccc}
    \toprule
     & & \multicolumn{3}{c|}{$\gamma=0$, $W=$} & \multicolumn{3}{c}{$\gamma=0.5$, $W=$} \\
    $\alpha$ & $\beta$ & 1 & 2 & 4 & 1 & 2 & 4\\
    \midrule
    0.001 & 0.05 & 0.002 & 0.001 & 0.001 & 0.005 & 0.004 & 0.003\\
    0.001 & 0.1 & 0.004 & 0.003 & 0.003 & 0.010 & 0.007 & 0.006\\
    0.01 & 0.05 & 0.022 & 0.016 & 0.014 & 0.056 & 0.038 & 0.031\\
    0.01 & 0.1 & 0.045 & 0.033 & 0.029 & 0.105 & 0.074 & 0.061\\
    0.05 & 0.05 & 0.114 & 0.083 & 0.074 & 0.280 & 0.195 & 0.161\\
    0.05 & 0.1 & 0.238 & 0.177 & 0.156 & 0.523 & 0.376 & 0.314\\
    \bottomrule
    \end{tabular}
    \caption{Optimal relative extra reward $\rho_A/\alpha-1$ (\%) under T-PAW for $W$-PPLNS with different window sizes $W$.}
    \label{tab::pplns_W}
\end{table}

\begin{table}[!tp]
    \centering
    \resizebox{\columnwidth}{!}{%
    \begin{tabular}{cc|c|cc|cccc|cccc}
    \toprule
     & & PPS & \multicolumn{2}{c|}{PAH} & \multicolumn{4}{c|}{T-PAW, $\gamma=0$} & \multicolumn{4}{c}{T-PAW, $\gamma=0.5$} \\
    $\alpha$ & $\beta$ & AoA & PROP & EXP & PPLNS & PROP & EXP & GRP & PPLNS & PROP & EXP & GRP\\
    \midrule
    0.001 & 0.05 & 0.100 & 27.790 & 2.184 & 0.001 & 0.005 & 0.108 & 0.074 & 0.004 & 0.018 & 0.169 & 0.111\\
    0.001 & 0.1 & 0.100 & 27.968 & 4.010 & 0.003 & 0.006 & 0.061 & 0.040 & 0.007 & 0.020 & 0.092 & 0.059\\
    0.01 & 0.05 & 1.010 & 24.979 & 2.009 & 0.016 & 0.051 & 0.890 & 0.616 & 0.038 & 0.163 & 1.366 & 0.906\\
    0.01 & 0.1 & 1.010 & 26.454 & 3.837 & 0.033 & 0.071 & 0.558 & 0.367 & 0.074 & 0.190 & 0.836 & 0.534\\
    0.05 & 0.05 & 5.263 & 17.603 & 1.523 & 0.083 & 0.194 & 2.382 & 1.698 & 0.195 & 0.544 & 3.425 & 2.385\\
    0.05 & 0.1 & 5.263 & 21.496 & 3.252 & 0.177 & 0.298 & 1.955 & 1.314 & 0.376 & 0.724 & 2.800 & 1.849\\
    \bottomrule
    \end{tabular}}
    \caption{Optimal relative extra reward $\rho_A/\alpha-1$ (\%) under PAH and T-PAW for every block-dependent payout scheme, with $W=2$ for PPLNS and $\lambda_d=2\lambda_1$ for EXP. For comparison, the PPS column gives RER under the All-out Attack~\cite{doger2026alloutattackoptimalblock}, i.e., $\frac{\alpha}{1-\alpha}$, which does not depend on $\beta$ or $\gamma$.}
    \label{tab::schemes}
\end{table}

\section{Numerical Results}\label{sec::numerical}
In this section, we compare the vulnerability of the payout schemes to T-PAW and PAH. For every $(\alpha,\beta,\gamma)$, we maximize the adversarial revenue ratio $\rho_A$ over $(p_1,p_2,p_3,T)$ for T-PAW and over $(p_y,p_o,t^*)$ for PAH, using the closed-form expressions of Sections~\ref{sec::withholding} and~\ref{sec::hop} with a multi-start interior-point method, and report the relative extra reward at the optimum. We verify all analytical expressions with Monte Carlo simulations that model the attack cycles, the pool rounds, the PPLNS window and the EXP scores explicitly, without using any of the derivations. At the optimal parameters, the simulated revenue ratios, block redundancy ratios and reward fractions match the theory, with the reward fractions agreeing to the fourth digit for every scheme; see Appendix~\ref{app::mc}. We set $\lambda_1=1$, which is without loss of generality since the results depend on $T$ and $t^*$ only through $\lambda_1T$ and $\lambda_1t^*$, and we assume a rational pool manager. Unless stated otherwise, we use $W=2$ for PPLNS and $\lambda_d=2\lambda_1$ for EXP, as in the original slush method. Note that PAH does not withhold any block, thus it has $\delta=1$ and its gain does not rely on the difficulty adjustment, whereas the gain of T-PAW is mostly realized after the difficulty adjustment, since withholding wastes part of the network hashpower; the pre-adjustment gain reported in~\cite{doger2026temporarypoweradjustingwithholding} is obtained under GRP, which overstates it.

\emph{GRP is not a reliable proxy:} Table~\ref{tab::schemes} shows that the T-PAW gain under GRP, which is the scheme adopted by prior analyses~\cite{power_adjusting,doger2026temporarypoweradjustingwithholding}, differs substantially from the gain under the schemes used in practice, and the direction of the error depends on the scheme. Under round-based PROP, the blockwork carried over from earlier attack cycles dilutes the effect of the power adjustments, and T-PAW yields up to an order of magnitude less than under GRP, e.g., $0.194\%$ versus $1.698\%$ for $(\alpha,\beta,\gamma)=(0.05,0.05,0)$. Under EXP, the decay emphasizes the recent shares of the withholding phase, and T-PAW yields more than under GRP, e.g., $2.382\%$ versus $1.698\%$ for the same parameters. Hence, the conclusions drawn under GRP carry over to neither PROP nor EXP.

\emph{Which attack dominates under each scheme:} Under PROP, PAH dominates by far: its gain is between $17.603\%$ and $27.968\%$, whereas the gain of T-PAW stays below $1\%$. For a small adversary, the PAH gain approaches the $28\%$ of a negligible miner~\cite{raulo2011optimal,rosenfeld2011analysisbitcoinpooledmining}, e.g., $27.790\%$ for $(\alpha,\beta)=(0.001,0.05)$. As $\alpha/\beta$ grows, the adversary dilutes the young rounds with its own shares and the gain decreases, e.g., to $17.603\%$ for $\alpha=\beta=0.05$. Under EXP, both attacks are effective and the best attack depends on the parameters: PAH is preferable for a small adversary or a large pool, e.g., $3.837\%$ versus at most $0.836\%$ for $(\alpha,\beta)=(0.01,0.1)$, whereas T-PAW is preferable for a large adversary relative to the pool, e.g., $2.382\%$ versus $1.523\%$ for $(\alpha,\beta,\gamma)=(0.05,0.05,0)$. Under PPLNS, PAH yields no gain by Proposition~\ref{prop::hop_pplns}, and the gain of T-PAW is the smallest among all schemes, at most $0.376\%$ in Table~\ref{tab::schemes}. However, it remains positive for every $(\alpha,\beta,\gamma)$, thus PPLNS mitigates T-PAW but does not eliminate it. In all schemes, the T-PAW gain increases with $\gamma$, as the released fPoWs win more fork races, while PAH does not depend on $\gamma$.

\emph{Comparison with PPS:} Under PPS, each share is paid immediately and its payment does not depend on the rounds, therefore PAH yields no gain, whereas BWH is profitable after the difficulty adjustment, where the optimal strategy is the All-out Attack (AoA) with a gain of $\frac{\alpha}{1-\alpha}$ per epoch that is borne by the pool operator~\cite{doger2026alloutattackoptimalblock}. For $\alpha=0.05$, the AoA gain under PPS exceeds the gain of every attack under every block-dependent scheme except PAH under PROP. For a small adversary, however, block-dependent schemes are not safer: T-PAW under EXP already exceeds the PPS gain, e.g., $1.366\%$ versus $1.010\%$ for $(\alpha,\beta,\gamma)=(0.01,0.05,0.5)$, and PAH under PROP and EXP exceeds it by far, since the relative gain of PAH barely decreases as $\alpha\to0$, while the AoA gain vanishes linearly in $\alpha$. Among all schemes we consider, PPLNS is the least vulnerable for every $(\alpha,\beta,\gamma)$.

\emph{Decay rate of EXP:} Table~\ref{tab::exp_lambda} reveals a trade-off in the design of EXP. As $\lambda_d$ increases, the scores of the shares submitted before the withholding phase decay faster, hence the power adjustments of T-PAW weigh more, and the T-PAW gain increases, e.g., from $1.438\%$ to $3.579\%$ for $(\alpha,\beta,\gamma)=(0.05,0.05,0)$ as $\lambda_d$ increases from $\lambda_1$ to $4\lambda_1$. At the same time, the score of an old round saturates faster, thus the value of a share depends less on the blockwork age of the round, and the PAH gain decreases, e.g., from $2.768\%$ to $0.805\%$ for the same parameters. In the limit $\lambda_d\to0$, EXP reduces to PROP, which is the most vulnerable scheme to PAH. EXP cannot protect against both attacks by tuning $\lambda_d$: a slow decay invites PAH and a fast one invites T-PAW.

\emph{Window of PPLNS:} Table~\ref{tab::pplns_W} shows that the T-PAW gain under PPLNS decreases with the window size $W$, e.g., from $0.523\%$ to $0.314\%$ for $(\alpha,\beta,\gamma)=(0.05,0.1,0.5)$ as $W$ increases from $1$ to $4$. By Theorem~\ref{thm:closed_pplns}, the power adjustments of T-PAW affect the reward fractions only through terms of order $1/W$, since the blockwork of a single attack cycle is a small part of the window. The part of the gain that stems from the fork races, as in FAW, does not depend on $W$, therefore the gain does not vanish as $W$ grows. Hence, a longer window alone cannot make PPLNS fully resistant to T-PAW.

\section{Conclusion}
We analyzed power adjusting attacks under the block-dependent payout schemes used in practice. For T-PAW, which contains FAW and PAW as special cases, we derived the revenue ratios under $W$-PPLNS, round-based PROP and EXP, and we showed that GRP, which is adopted by prior analyses, is not a reliable proxy, since it overstates the gain under PROP and understates it under EXP. We further introduced PAH, which requires neither withholding nor a difficulty adjustment, and derived its revenue ratios for an adversary of any size. PROP is highly vulnerable to PAH, and EXP cannot be tuned against both attacks, as a slow decay invites PAH and a fast decay invites T-PAW. PPLNS is immune to PAH and yields the smallest T-PAW gain among all schemes, including the All-out Attack (AoA) under PPS. We leave strategies that combine withholding and hopping to the future work and believe that such a strategy might prove to be especially useful against score-based schemes. Another promising direction is analyzing the proposed attacks under multiple target pool systems which we believe can increase the rewards especially for PAH.

\appendices

\section{Reward Fractions for $W$-PPLNS}\label{app::pplns_scheme}
We first consider the blockwork of the pool accumulated in an attack cycle. Note that blockwork refers to the expected number of fPoWs generated:
\begin{enumerate}
    \item \emph{First phase:} Let $\hQ_1$ denote the blockwork of a pool in the first phase of an attack cycle which lasts $T_1\sim\mathrm{Exp}(\lambda_1)$, where the pool works with hashwork $\lambda_1a_1$. Hence, we have $\hQ_1\triangleq\lambda_1a_1T_1\sim\mathrm{Exp}(1/a_1)$, with $\E[\hQ_1]=a_1$ and $\E[\hQ_1^2]=2a_1^2$.
    \item \emph{Withholding phase:}
        \begin{enumerate}
        \item \emph{Successful (Case~4a}): With probability $\theta$ the withholding deadline is reached and we let $\hT$ denote the blockwork of the pool. This withholding phase lasts $T$, where the pool works with hashwork $\lambda_1a_2$. Hence, we have $\hT\triangleq\lambda_1a_2T$ which is deterministic.
        \item \emph{Unsuccessful (Case~4b}): With probability $1-\theta$ the phase ends earlier after the time $T_2'$, and the blockwork of the pool is $\hQ_2\triangleq\lambda_1a_2T_2'$. Since $T_2'\sim (T_2\mid T_2<T)$ with $T_2\sim\mathrm{Exp}(\lambda_2)$, we have $\hQ_2\sim (J_2\mid J_2<\hT)$ with $J_2\sim\mathrm{Exp}(\mu_2)$.
    \end{enumerate} 
    \item \emph{Fork race phase:} When the withholding phase is ended before the deadline by a block of the honest miners outside the pool, a fork race of duration $T_3\sim\mathrm{Exp}(\lambda_1)$ follows with the pool at hash fraction $a_3$, hence the blockwork of the pool is $\hQ_3\triangleq\lambda_1a_3T_3\sim\mathrm{Exp}(1/a_3)$, with $\E[\hQ_3]=a_3$ and $\E[\hQ_3^2]=2a_3^2$.
\end{enumerate}
It is straightforward to verify that
\begin{align}
    \nu_1&\triangleq\E[\hQ_2]=\frac{1}{\mu_2}-\frac{\theta\hT}{1-\theta},\label{eq:nu1}\\
    \nu_2&\triangleq\E[\hQ_2^2]=\frac{2}{\mu_2^2}-\frac{\theta\hT}{1-\theta}\left(\hT+\frac{2}{\mu_2}\right),
    \label{eq:nu2}
\end{align}

\begin{table*}[h]
    \centering\small\setlength{\tabcolsep}{4.5pt}
    \begin{tabular}{llll}
    \toprule
    Events & Probability & Pool blockwork $S$ & Adversarial pool blockwork $X$\\
    \midrule
    Cases~1,2,3 & $1-\alpha p_1$ & $\hQ_1$ & $a_1'\hQ_1$\\
    Case~4a & $\alpha p_1\theta$ & $\hQ_1+\hT$ & $a_1'\hQ_1+a_2'\hT$\\
    Case~4b-i,4b-ii & $\alpha p_1(1-\theta)(\pi_A+\pi_P)$ & $\hQ_1+\hQ_2$ & $a_1'\hQ_1+a_2'\hQ_2$\\
    Case~4b-iii & $\alpha p_1(1-\theta)\pi_R$ & $\hQ_1+\hQ_2+\hQ_3$ & $a_1'\hQ_1+a_2'\hQ_2+a_3'\hQ_3$\\
    \bottomrule
    \end{tabular}
    \caption{Cumulative pool and adversarial pool blockworks $S,X$ in an attack cycle.}
    \label{tab:cases}
\end{table*}

Let $S=\hQ_1+V$ denote the cumulative blockwork of the pool at the end of an attack cycle. Table~\ref{tab:cases} lists the events of an attack cycle, their probabilities, the pool blockwork $S$ of the cycle and its adversarial part $X$. We have
\begin{align}
    V=\begin{cases}
    0, & \text{w.p. } 1-\alpha p_1,\\
    \hT, & \text{w.p. } \alpha p_1\theta,\\
    \hQ_2, & \text{w.p. } \alpha p_1(1-\theta)(\pi_A+\pi_P),\\
    \hQ_2+\hQ_3, & \text{w.p. } \alpha p_1(1-\theta)\pi_R.
    \end{cases}
\end{align}
Using $\pi_A+\pi_P+\pi_R=1$ and $\E[(\hQ_2+\hQ_3)^2]=\nu_2+2a_3\nu_1+2a_3^2$,
\begin{align}
    \E[V]&=\alpha p_1\left\{\theta\hT+(1-\theta)\left(\nu_1+\pi_Ra_3\right)\right\},\label{eq:EV}\\
    \E[V^2]&=\alpha p_1\left\{\theta\hT^2+(1-\theta)\left(\nu_2+2\pi_Ra_3\nu_1+2\pi_Ra_3^2\right)\right\}.\label{eq:EV2}
\end{align}
The cumulative blockwork of the pool at the end of an attack cycle is $S=\hQ_1+V$ with $\hQ_1$ independent of $V$, thus
\begin{align}
    \E[S]=a_1+\E[V],\
    \E[S^2]=2a_1^2+2a_1\E[V]+\E[V^2].\label{eq:ES}
\end{align}
Here, explicitly
\begin{align}
    \E[S]=a_1+\alpha p_1(1-\theta)\frac{a_2+a_3\xi}{1-\alpha p_2}.\label{eq:ESexp}
\end{align}

Let $b(t)=\sigma$ denote the cumulative blockwork of the target pool since the beginning until time $t$. Under $W$-PPLNS the reward of a pool block that the manager receives at $t$ is split according to the blockworks submitted during $(b^{-1}(\sigma-W),b^{-1}(\sigma)]$. Let $x(\sigma)$ be the adversarial fraction of the blockwork submitted within an infinitesimal neighborhood of  $b^{-1}(\sigma)$. In a T-PAW cycle, we have
\begin{align}
    x(\sigma)=\begin{cases}
    a_1', & \text{first phase},\\
    a_2', & \text{withholding phase },\\
    a_3', & \text{fork race phase}.
    \end{cases}
    \label{eq:x}
\end{align}

Hence, among the blockwork submitted during $(b^{-1}(\sigma-W),b^{-1}(\sigma)]$, the adversarial blockwork is
\begin{align}
    \frac{1}{W}\int_{\sigma-W}^{\sigma}x(s)\,ds. \label{eq:window}
\end{align}
Let $X=\int_0^Sx(\sigma)d\sigma$ be the adversarial pool blockwork in an attack cycle. It is straightforward to verify that
\begin{align}
    \E[X]&=\alpha p_1+\alpha p_1a_2'\left\{\theta\hT+(1-\theta)\nu_1\right\}+\alpha p_1(1-\theta)\pi_R\alpha p_3\nonumber\\
    &=\alpha p_1\Big(1+(1-\theta)\frac{\alpha p_2+\alpha p_3\xi}{1-\alpha p_2}\Big).\label{eq:EX}
\end{align}
Thus, the long-run adversarial blockwork fraction is the ratio of \eqref{eq:EX} and
\eqref{eq:ESexp},
\begin{align}
    \xb\triangleq\frac{\E[X]}{\E[S]}
    =\frac{\alpha p_1\left[1-\alpha p_2+(1-\theta)\big(\alpha p_2+\alpha p_3\xi\big)\right]}
    {a_1(1-\alpha p_2)+\alpha p_1(1-\theta)\left[a_2+a_3\xi\right]}.\label{eq:xbar}
\end{align}

Next, we find $\E[\Psi]\triangleq\E[\int_0^Sx(\sigma)(S-\sigma)\,d\sigma]$ by splitting the integral over the three phases.
\begin{itemize}[leftmargin=*]
    \item First phase, $\sigma\in[0,\hQ_1]$, $x=a_1'$, $S-\sigma=\hQ_1-\sigma+V$, $\ a_1'\int_0^{\hQ_1}(\hQ_1-\sigma+V)d\sigma=a_1'\left(\frac{\hQ_1^2}{2}+\hQ_1V\right)$, with expectation $a_1'\left(a_1^2+a_1\E[V]\right)=\alpha p_1\left(a_1+\E[V]\right)$ since $\hQ_1$ and $V$ are independent.
    \item Withholding phase of pool work $L$ followed by pool work $R$ until the end of the cycle, $x=a_2'$: $\ a_2'\int_0^{L}(L-s+R)ds=a_2'\left(\frac{L^2}{2}+LR\right)$. In Case~4a, $L=\hT$ and $R=0$. In Case~4b, $L=\hQ_2$, and $R=\hQ_3$ with probability $\pi_R$ and $R=0$ otherwise, thus $\E[L^2/2+LR]=\nu_2/2+\pi_Ra_3\nu_1$.
    \item Fork race phase, $x=a_3'$, and the cycle ends with the race: $\ a_3'\int_0^{\hQ_3}(\hQ_3-s)ds=a_3'\hQ_3^2/2$, with expectation $a_3'a_3^2=\alpha p_3\,a_3$.
\end{itemize}
With the probabilities of each phase happening,
\begin{align}
    \E[\Psi]=&\ \alpha p_1\left(a_1+\E[V]\right)+\alpha p_1(1-\theta)\pi_R\,\alpha p_3\,a_3
    \nonumber\\
    &+\alpha p_1a_2'\left\{\frac{\theta\hT^2}{2}+(1-\theta)\left(\frac{\nu_2}{2}+\pi_Ra_3\nu_1\right)\right\}.\label{eq:EPsi}
\end{align}

\subsection{Adversarial Blockwork in the Window Before a Cycle}
Let $H(v)$ be the expected adversarial pool blockwork in a blockwork window $v$ that immediately precedes the start of an attack cycle. Throughout this section $m_1\triangleq\E[S]$, $m_2\triangleq\E[S^2]$ and $F$ is the distribution function of $S$.

\begin{lemma}\label{lem:H}
    $H$ satisfies the renewal equation
    \begin{align}
        H(v)=\E\left[\int_{(S-v)^+}^{S}x(\sigma)\,d\sigma\right]+\E\big[H(v-S)\,\ind{S<v}\big],\label{eq:Hrenewal}
    \end{align}
    and, as $v\to\infty$,
    \begin{align}
        H(v)=\xb\,v+d+o(1),\qquad d\triangleq\frac{\xb\,\E[S^2]}{2\,\E[S]}-\frac{\E[\Psi]}{\E[S]}.\label{eq:d}
    \end{align}
\end{lemma}

\begin{Proof}
\emph{Step 1, the renewal equation:} Attack cycles are i.i.d., hence the cycles that precede a given cycle start, read backwards, are i.i.d.~copies of $(S,x(\cdot))$ and are independent of the cycle that starts there. Consider the last cycle before the start and let $S$ be its pool blockwork. If $S\geq v$, the pool blockwork window $v$ lies inside this cycle and covers its positions $(S-v,S]$. If $S<v$, the pool blockwork window contains the whole cycle, and the remaining pool blockwork window $v-S$ ends at the start of that cycle, which is again a cycle start, thus it holds $H(v-S)$ pool blockwork in expectation. Both cases are written in \eqref{eq:Hrenewal}. With $h_0(v)\triangleq\E\big[\int_{(S-v)^+}^{S}x(\sigma)d\sigma\big]$, \eqref{eq:Hrenewal} gives
\begin{align}
    H(v)=h_0(v)+\int_0^vH(v-s)\,dF(s)\label{eq:Hconv},
\end{align}
and
\begin{align}
    H(v)=\int_{[0,v]}h_0(v-y)\,dU(y),\ U(y)\triangleq\sum_{n=0}^{\infty}F^{*n}(y).\label{eq:Hsol}
\end{align}
$U$ is the renewal function and $U(y)$ is the expected number of cycle starts in a pool blockwork window $v$ before the given cycle start, counting the given one ($F^{*0}=\ind{y\geq0}$). Next, let $\hat h(v)$ be the expected adversarial pool blockwork of an attack cycle that lies more than $v$ before its end,
\begin{align}
    h_0(v)=\E[X]-\hat h(v),\ \hat h(v)\triangleq\E\bigg[\int_0^{(S-v)^+}x(\sigma)\,d\sigma\bigg].\label{eq:hhat}
\end{align}
Inserting in \eqref{eq:Hsol},
\begin{align}
    H(v)=\underbrace{\E[X]\,U(v)}_{\text{(I)}}-\underbrace{\int_{[0,v]}\hat h(v-y)\,dU(y)}_{\text{(II)}}.\label{eq:HI-II}
\end{align}
Term (I) counts all adversarial pool blockwork of every cycle that has its later boundary inside the interval. The earliest of these cycles extends beyond the interval, and term (II) removes its shares that lie outside. We will later use the following fact (see, \cite{asmussen2003applied}): If $F$ is non-lattice with mean $m_1$ and $z\geq0$ is non-increasing with $\int_0^\infty z(t)dt<\infty$, then
\begin{align}
    \int_{[0,v]}z(v-y)\,dU(y)\ \xrightarrow{v\to\infty}\ \frac{1}{m_1}\int_0^\infty z(t)\,dt. \label{eq:KRT}
\end{align}

\emph{Step 2, term (I):} Let $g(v)\triangleq U(v)-v/m_1$. Note $U=1+U*F$ and substitute $U(v)=g(v)+v/m_1$ on both sides,
\begin{align}
    g(v)=&\underbrace{1-\frac{v}{m_1}+\frac{1}{m_1}\int_0^v(v-s)\,dF(s)}_{\triangleq\, z(v)} \nonumber\\
    &+\int_0^vg(v-s)\,dF(s).
\end{align}
By exchanging the order of integration, $\int_0^v(v-s)\,dF(s)=\int_0^v\int_s^vdt\,dF(s)=\int_0^vF(t)\,dt$, hence
\begin{align}
    z(v)=1-\frac{1}{m_1}\int_0^v\big(1-F(t)\big)dt=\frac{1}{m_1}\int_v^\infty\big(1-F(t)\big)dt.
\end{align}
Thus, $z\geq0$ is non-increasing, and
\begin{align}
    \int_0^\infty z(v)\,dv&=\frac{1}{m_1}\int_0^\infty\int_v^\infty\big(1-F(t)\big)dt\,dv\\
&=\frac{1}{m_1}\int_0^\infty t\big(1-F(t)\big)dt=\frac{m_2}{2m_1}.
\end{align}
The renewal equation $g=z+g*F$ has the solution $g=z*U$, and \eqref{eq:KRT} gives $g(v)\to m_2/(2m_1^2)$, i.e.,
\begin{align}
    U(v)=\frac{v}{m_1}+\frac{m_2}{2m_1^2}+o(1).\label{eq:Uasym}
\end{align}
\emph{Step 3, term (II):} The function $\hat h$ in \eqref{eq:hhat} is non-negative and non-increasing in $v$. A pool blockwork at position $\sigma$ of a cycle is counted in $\hat h(v)$ exactly when $\sigma<S-v$, i.e., for $v\in[0,S-\sigma)$, hence by exchanging the order of integration
\begin{align}
    \!\int_0^\infty\hat h(v)\,dv&=\E\bigg[\int_0^Sx(\sigma)\int_0^\infty\ind{v<S-\sigma}\,dv\,d\sigma\bigg]\\
    &=\E\bigg[\int_0^Sx(\sigma)(S-\sigma)\,d\sigma\bigg]=\E[\Psi]<\infty.
\end{align}
Hence \eqref{eq:KRT} applies with $z=\hat h$ and term (II) converges to $\E[\Psi]/m_1$. Combining (I) and (II) completes the proof.
\end{Proof}

\subsection{Reward Fractions of the Adversary}

\begin{theorem}\label{thm:exact}
    If a canonical pool block received when the cumulative blockwork since time immemorial is $\sigma>W$, with
    \begin{align}
        \mathcal{A}(\hq_1,\hq_2,\hq_3)\triangleq&\ a_3'\min\{\hq_3,W\}+a_2'\min\{\hq_2,(W-\hq_3)^+\}\nonumber\\
        &+a_1'\min\{\hq_1,(W-\hq_2-\hq_3)^+\}\nonumber\\
        &+H\big((W-\hq_1-\hq_2-\hq_3)^+\big),\label{eq:calA}
    \end{align}
    the expected adversarial shares under $W$-PPLNS are
    \begin{align}
        r_1&=\frac{\E[\mathcal{A}(\hQ_1,0,0)]}{W},\\
        r_s&=\frac{\E[\mathcal{A}(\hQ_1,\hT,0)]}{W},\\
        r_u&=\frac{\E[\mathcal{A}(\hQ_1,\hQ_2,0)]}{W},\\
        r_f&=\frac{\E[\mathcal{A}(\hQ_1,\hQ_2,\hQ_3)]}{W}.\label{eq:exactr}
    \end{align}
\end{theorem}

\begin{Proof}
Consider the window backwards from the canonical pool block. For example, consider the pool block of a fork race. The window $W$ first meets the race phase (pool work $\hq_3$, density $a_3'$), of which it holds $a_3'\min\{\hq_3,W\}$, then the withholding phase ($\hq_2$, density $a_2'$), then the first phase ($\hq_1$, density $a_1'$), and then earlier cycles. The length of the window left for the withholding phase is $(W-\hq_3)^+$, thus it holds $a_2'\min\{\hq_2,(W-\hq_3)^+\}$; the length left for the first phase is $(W-\hq_2-\hq_3)^+$, thus it holds $a_1'\min\{\hq_1,(W-\hq_2-\hq_3)^+\}$; the length left for earlier cycles is $(W-\hq_1-\hq_2-\hq_3)^+$. This is \eqref{eq:calA}. The other canonical pool blocks, i.e., those of Cases~3, 4a and 4b-ii and the released fPoW of Case~4b-iii, are similar with $\hq_3=0$ and $\hq_2\in\{0,\hT,\hQ_2\}$. Since the cycle ends at the first of independent exponential clocks, $\hQ_1,\hQ_2,\hQ_3$ are independent of which case occurs, and taking expectations completes the proof.
\end{Proof}

Combining the result of Lemma~\ref{lem:H} with Theorem~\ref{thm:exact}, we can prove Theorem~\ref{thm:closed_pplns} with
\begin{align}
    r_1:\quad \E\big[&a_1'\hQ_1-\xb\hQ_1\big]=(a_1'-\xb)a_1,\\
    r_s:\quad \E\big[&a_1'\hQ_1+a_2'\hT-\xb(\hQ_1+\hT)\big]\nonumber\\
    &=(a_1'-\xb)a_1+(a_2'-\xb)\hT,\\
    r_u:\quad \E\big[&a_1'\hQ_1+a_2'\hQ_2-\xb(\hQ_1+\hQ_2)\big]\nonumber\\
    &=(a_1'-\xb)a_1+(a_2'-\xb)\nu_1,\\
    r_f:\quad \E\big[&a_1'\hQ_1+a_2'\hQ_2+a_3'\hQ_3-\xb(\hQ_1+\hQ_2+\hQ_3)\big]\\
    &=(a_1'-\xb)a_1+(a_2'-\xb)\nu_1+(a_3'-\xb)a_3.
\end{align}
The closed-forms neglect the $o(1)$ term of \eqref{eq:d}, evaluated at $v=W-\hQ_1-\hQ_2-\hQ_3$, together with the event $\hQ_1+\hQ_2+\hQ_3>W$. Since the pool blockwork of a cycle is a fraction of one unit on average and has exponentially decaying tails, both are negligible for windows $W$ of a few units of blockwork. If $W$ is comparable to the pool blockwork of a cycle, Theorem~\ref{thm:exact} must be used directly.

\section{Reward Fractions for Round-Based PROP}\label{app::prop_scheme}
 
\subsection{Model}
Under proportional (PROP) reward sharing, the pool manager splits the reward of a pool block among the members in proportion to the blockwork they submitted during the \emph{round} of that block. We define the rounds by the canonical pool blocks as follows:
\begin{itemize}
    \item[(R1)] The round boundaries are the instants at which the manager receives canonical pool blocks. The reward of a canonical pool block is split in proportion to the blockwork submitted since the previous round boundary.
    \item[(R2)] A pool block that is orphaned, and a withheld fPoW that is discarded, is not a round boundary. The blockwork submitted before and after it belongs to the same round.
\end{itemize}
A round is therefore not tied to an attack cycle and ends only with a canonical block of the pool. In particular, if the adversary releases a withheld fPoW to start a fork race, the release time is a round boundary if and only if the released fPoW wins the race to become canonical. 
 
\begin{remark}
    A pool could alternatively close a round whenever it receives a fPoW, including a released fPoW that later loses the fork race. We believe our definition above is more likely to be used in practice. It is straightforward (and easier) to modify the analysis given here to obtain the reward fractions for the alternative case. 
\end{remark}
 
The adversarial fraction of the pool blockwork is $a_1'$, $a_2'$, $a_3'$ in the first phase, the withholding phase and the fork race phase, respectively, and we refer to these regimes as modes $1,2,3$. If a round consists of multiple attack cycles, with total pool blockwork $y_i$ submitted in mode $i$ during the entire round, the adversarial fraction of the pool blockwork is
\begin{align}
    \varrho(\mathbf{y})&=\frac{a_1'y_1+a_2'y_2+a_3'y_3}{y_1+y_2+y_3} \\
    &=a_1'+\frac{(a_2'-a_1')y_2}{y_1+y_2+y_3}+\frac{(a_3'-a_1')y_3}{y_1+y_2+y_3}.\label{eq::round_fraction}
\end{align}
 
\subsection{The Carry Process}
The attack cycles remain i.i.d., but the share rewards of the pool do not: the reward split of a pool block depends on the blockwork of its round, which may span multiple attack cycles. Let us index the attack cycles by $n=0,1,2,\dots$ and let $\mathbf{C}^{(n)}=(C^{(n)}_1,C^{(n)}_2,C^{(n)}_3)$ be the \emph{carry}, i.e., the pool blockwork of the open round submitted in each mode, observed at the start of cycle $n$. Let $\zeta_n$ collect the randomness of cycle $n$. Define
\begin{align}
    \varepsilon&=\alpha p_1\pi_R(1-\gamma)\xi,\label{eq::eps}\\
    \eta_3&=\alpha p_1(1-\theta)\pi_R(\alpha q_3+\gamma\xi),\label{eq::eta3}\\
    \eta_0&=\beta+\alpha p_1\big(\theta+(1-\theta)(\pi_P+\pi_Ra_3)\big).\label{eq::eta0}
\end{align}

We consider how an attack cycle $n$ that starts with carry $\mathbf{C}^{(n)}=\mathbf{C}$ evolves and list the probability and the blockwork associated with the canonical pool blocks mined during that cycle as well as the carry $\mathbf{C}^{(n+1)}=\mathbf{C}^+$ left to the next cycle:
\begin{itemize}[leftmargin=*]
    \item[] Case~1, w.p.\ $\xi$ no pool block. $\mathbf{C}^+=\mathbf{C}+(\hQ_1,0,0)$.
    \item[] Case~2, w.p.\ $\alpha q_1$ no pool block. $\mathbf{C}^+=\mathbf{C}+(\hQ_1,0,0)$.
    \item[] Case~3, w.p.\ $\beta$: a canonical pool block closes the round with blockwork $\mathbf{C}+(\hQ_1,0,0)$. $\mathbf{C}^+=0$.
    \item[] Case~4a, w.p.\ $\alpha p_1\theta$: the fPoW released is canonical and closes the round with blockwork $\mathbf{C}+(\hQ_1,\hT,0)$. $\mathbf{C}^+=0$.
    \item[] Case~4b-i, w.p.\ $\alpha p_1(1-\theta)\pi_A$: the fPoW is discarded, no canonical pool block. $\mathbf{C}^+=\mathbf{C}+(\hQ_1,\hQ_2,0)$.
    \item[] Case~4b-ii, w.p.\ $\alpha p_1(1-\theta)\pi_P$: a canonical pool block closes the round with blockwork $\mathbf{C}+(\hQ_1,\hQ_2,0)$. $\mathbf{C}^+=0$.
    \item[] Case~4b-iii, w.p.\ $\alpha p_1(1-\theta)\pi_R$: the fPoW is released and the race is decided by next block.
    \begin{itemize}[leftmargin=*]
        \item[] Case~4b-iii-A, w.p.\ $\alpha q_3$: the round closes with blockwork $\mathbf{C}+(\hQ_1,\hQ_2,0)$ and $\mathbf{C}^+=(0,0,\hQ_3)$.
        \item[] Case~4b-iii-B, w.p.\ $a_3$:
        \begin{itemize}[leftmargin=*]
            \item[] $\mathds{1}_{P_R}=1$: the round closes with blockwork $\mathbf{C}+(\hQ_1,\hQ_2,0)$ and the subsequent round is also closed with the canonical pool block of the race  which has blockwork $(0,0,\hQ_3)$. $\mathbf{C}^+=0$.
            \item[] $\mathds{1}_{P_R}=0$: the fPoW is orphaned, but the pool block of the race is canonical and closes the round with blockwork $\mathbf{C}+(\hQ_1,\hQ_2,\hQ_3)$. $\mathbf{C}^+=0$.
        \end{itemize}
        \item[] Case~4b-iii-C, w.p.\ $\xi$:
        \begin{itemize}[leftmargin=*]
            \item[] w.p.\ $\gamma$: the round closes with blockwork $\mathbf{C}+(\hQ_1,\hQ_2,0)$. $\mathbf{C}^+=(0,0,\hQ_3)$.
            \item[] w.p.\ $(1-\gamma)$: the fPoW is orphaned. $\mathbf{C}^+=\mathbf{C}+(\hQ_1,\hQ_2,\hQ_3)$.
        \end{itemize}
    \end{itemize}
\end{itemize}
Consequently,
\begin{align}
    \mathbf{C}^{+}=\begin{cases}
    \mathbf{C}+(\hQ_1,0,0), & \text{w.p. } \xi+\alpha q_1,\\
    \mathbf{C}+(\hQ_1,\hQ_2,0), & \text{w.p. } \alpha p_1(1-\theta)\pi_A,\\
    \mathbf{C}+(\hQ_1,\hQ_2,\hQ_3), & \text{w.p. } (1-\theta)\varepsilon,\\
    (0,0,\hQ_3), & \text{w.p. } \eta_3,\\
    (0,0,0), & \text{w.p. } \eta_0,
    \end{cases}\label{eq::carry_recursion}
\end{align}
and the five probabilities sum to one.

Note that $\mathbf{C}^{(n+1)}$ is a deterministic function of $(\mathbf{C}^{(n)},\zeta_n)$, thus the carry is a Markov chain. We call cycle $n$ a \emph{reset cycle} if it is one of Cases~3, 4a, 4b-ii, 4b-iii-B. Whether a cycle is a reset cycle is a function of $\zeta_n$ only, it has probability $\eta_0\geq\beta>0$, and $\mathbf{C}^{(n+1)}=0$ iff cycle $n$ is a reset cycle. Notice this Markov chain is positive recurrent and regenerative, and by the renewal reward theorem
\begin{align}
    \rho_x=\frac{\E_{\pi}[B_x]}{\E[B_C]},\qquad x\in\{A,P,R\},
\end{align}
where $\E_\pi$ is taken with the carry in its stationary distribution. 

Thus, the reward fractions are the expected adversarial fractions of the paid rounds, with $\mathbf{C}\sim\pi$ independent of $\hQ_1,\hQ_2,\hQ_3$:
\begin{align}
    r_1&=\E_\pi\big[\varrho\big(\mathbf{C}+(\hQ_1,0,0)\big)\big], \label{eq::r_def1}\\
    r_s&=\E_\pi\big[\varrho\big(\mathbf{C}+(\hQ_1,\hT,0)\big)\big],\label{eq::r_def2}\\
    r_u&=\E_\pi\big[\varrho\big(\mathbf{C}+(\hQ_1,\hQ_2,0)\big)\big],\label{eq::r_def3}\\
    r_m&=\E_\pi\big[\varrho\big(\mathbf{C}+(\hQ_1,\hQ_2,\hQ_3)\big)\big],\label{eq::r_def4}\\
    r_f&=\mathds{1}_{P_R}\,\E\big[\varrho\big((0,0,\hQ_3)\big)\big]+\overline{\mathds{1}_{P_R}}\,r_m.\label{eq::r_def5}
\end{align}

\subsection{Transform of the Carry}
Note that, for $s\geq0$, we have
\begin{align}
    \E\big[e^{-s\hQ_1}\big]&=\phi_1(s)=\frac{1}{1+a_1s},\\
    \E\big[e^{-s\hQ_3}\big]&=\phi_3(s)=\frac{1}{1+a_3s},\label{eq::phi13}\\
    (1-\theta)\,\E\big[e^{-s\hQ_2}\big]&=\phi_2(s)=\frac{\mu_2\big(1-\theta e^{-s\hT}\big)}{\mu_2+s},\label{eq::phi2}\\ (1-\theta)\,\E\big[\hQ_2e^{-s\hQ_2}\big]&=\hat\phi_2(s)=-\phi_2'(s)\nonumber\\&=\frac{\mu_2\Big(1-\theta e^{-s\hT}\big(1+(\mu_2+s)\hT\big)\Big)}{(\mu_2+s)^2}\label{eq::tr}
\end{align}
Moreover $\phi_2(0)=1-\theta$, $\hat\phi_2(0)=(1-\theta)\nu_1$, $\hat\phi_2\geq0$, $\phi_2$ is non-increasing, $-\phi_3'(s)=a_3\phi_3(s)^2$, and
\begin{align}
    \phi_2(s)\leq\frac{\mu_2}{\mu_2+s}\leq1,\qquad \hat\phi_2(s)\leq\frac{\mu_2}{(\mu_2+s)^2}.\label{eq::bounds_phi2}
\end{align}

\begin{lemma}\label{lem::carry}
    The stationary carry has the Laplace transform
    \begin{align}
        \Lambda(s_1,s_2,s_3)=\E_\pi\big[e^{-s_1C_1-s_2C_2-s_3C_3}\big]=\frac{N(s_3)}{D(s_1,s_2,s_3)},
    \end{align}
    where
    \begin{align}
        N(s)=&\eta_0+\eta_3\phi_3(s),\\
        D(s_1,s_2,s_3)=&1-\phi_1(s_1)\nonumber\\
        &\times\Big(\xi+\alpha q_1+\phi_2(s_2)\big(\alpha p_1\pi_A+\varepsilon\phi_3(s_3)\big)\Big).
    \end{align}
\end{lemma}

\begin{Proof}
Taking the transform of \eqref{eq::carry_recursion} in stationarity and using the independence of $\mathbf{C}$, $\hQ_1$, $\hQ_2$, $\hQ_3$ gives $\Lambda=\eta_0+\eta_3\phi_3(s_3)+\Lambda\big(1-D(s_1,s_2,s_3)\big)$.
\end{Proof}

\begin{lemma}\label{lem::ratio}
    Let $\mathbf{Y}=(Y_1,Y_2,Y_3)$ be a non-negative random vector with $Y_1+Y_2+Y_3>0$ a.s.\ and Laplace transform $M(s_1,s_2,s_3)$. Then,
    \begin{align}
        \E\bigg[\frac{Y_i}{Y_1+Y_2+Y_3}\bigg]=-\int_0^\infty\frac{\partial M}{\partial s_i}(s,s,s)\,ds.
    \end{align}
\end{lemma}

\begin{Proof}
Use $\frac{1}{y}=\int_0^\infty e^{-sy}ds$ and Tonelli's theorem.
\end{Proof}

\begin{lemma}[Transforms of the paid rounds]\label{lem::rounds}
    With $\mathbf{C}\sim\pi$ independent of $\hQ_1,\hQ_2,\hQ_3$, the Laplace transforms of the rounds in \eqref{eq::r_def1}--\eqref{eq::r_def4} are
    \begin{align}
        M_1&=\Lambda\,\phi_1(s_1), &&\text{for } \mathbf{C}+(\hQ_1,0,0),\\
        M_s&=\Lambda\,\phi_1(s_1)\,e^{-s_2\hT}, &&\text{for } \mathbf{C}+(\hQ_1,\hT,0),\\
        M_u&=\Lambda\,\phi_1(s_1)\,\frac{\phi_2(s_2)}{1-\theta}, &&\text{for } \mathbf{C}+(\hQ_1,\hQ_2,0),\\
        M_m&=\Lambda\,\phi_1(s_1)\,\frac{\phi_2(s_2)\phi_3(s_3)}{1-\theta}, &&\text{for } \mathbf{C}+(\hQ_1,\hQ_2,\hQ_3),
    \end{align}
    with $\Lambda=\Lambda(s_1,s_2,s_3)$, and each of these rounds is non-zero a.s.
\end{lemma}

On the diagonal $s_1=s_2=s_3=s$ we write $\phi_i=\phi_i(s)$, $\hat\phi_2=\hat\phi_2(s)$, $N=N(s)$, $D=D(s,s,s)$, and define
\begin{align}
    K(s)&=\frac{N\phi_1}{D},\label{eq::K}\\
    K_2(s)&=\big(\alpha p_1\pi_A+\varepsilon\phi_3\big)\,\hat\phi_2\,\frac{\phi_1K(s)}{D},\label{eq::K2}\\
    K_3(s)&=a_3\phi_3^2\,\frac{\phi_1}{D}\big(\eta_3+\varepsilon\phi_2K(s)\big).\label{eq::K3}
\end{align}
Since $\Lambda\phi_1=K$, differentiating $N/D$ in Lemma~\ref{lem::carry} and using $-\phi_2'=\hat\phi_2$ and $-\phi_3'=a_3\phi_3^2$ gives, on the diagonal,
\begin{align}
    -\frac{\partial(\Lambda\phi_1)}{\partial s_2}=K_2(s),\qquad -\frac{\partial(\Lambda\phi_1)}{\partial s_3}=K_3(s).\label{eq::dK}
\end{align}
By \eqref{eq::round_fraction} and Lemma~\ref{lem::ratio}, $r_x=a_1'+(a_2'-a_1')\,\omega_2^{x}+(a_3'-a_1')\,\omega_3^{x}$ with $\omega_i^x=-\int_0^\infty\frac{\partial M_x}{\partial s_i}(s,s,s)\,ds$, the expected fraction of the blockwork of the round submitted in mode $i$. Applying the product rule to the transforms of Lemma~\ref{lem::rounds} with \eqref{eq::dK}, gives
\begin{align}
        \omega_2^{1}&=\int_0^\infty K_2(s)\,ds,\label{eq::om1}\\
        \omega_3^{1}&=\int_0^\infty K_3(s)\,ds,\\
        \omega_2^{s}&=\int_0^\infty e^{-s\hT}\big(\hT K(s)+K_2(s)\big)ds, \\
        \omega_3^{s}&=\int_0^\infty e^{-s\hT}K_3(s)\,ds,\label{eq::oms}\\
        \omega_2^{u}&=\frac{1}{1-\theta}\int_0^\infty\big(\hat\phi_2K(s)+\phi_2K_2(s)\big)ds, \\
        \omega_3^{u}&=\frac{1}{1-\theta}\int_0^\infty\phi_2K_3(s)\,ds,\label{eq::omu}\\
        \omega_2^{m}&=\frac{1}{1-\theta}\int_0^\infty\phi_3\big(\hat\phi_2K(s)+\phi_2K_2(s)\big)ds,\label{eq::omm2}\\
        \omega_3^{m}&=\frac{1}{1-\theta}\int_0^\infty\phi_2\phi_3\big(a_3\phi_3K(s)+K_3(s)\big)ds.\label{eq::omm3}
    \end{align}
    Finally, $\E[\varrho((0,0,\hQ_3))]=a_3'$ in \eqref{eq::r_def5}, which concludes the proof of Theorem~\ref{thm::round_prop}.

\begin{remark}[GRP]\label{rem::GRP}
    Under GRP, every block that becomes known to the pool manager starts a new round, and every attack cycle starts right after such a block. Hence, the open round is empty at the start of every cycle, i.e., $\mathbf{C}\equiv0$ and $\Lambda\equiv1$ in Lemma~\ref{lem::carry}. A withheld fPoW is not known to the manager, thus the blockwork of the withholding phase stays in the round, and the rounds of the pool blocks of Cases~3, 4a and 4b-ii, and of the released fPoW of Case~4b-iii, are $(\hQ_1,0,0)$, $(\hQ_1,\hT,0)$ and $(\hQ_1,\hQ_2,0)$ as in \eqref{eq::r_def1}--\eqref{eq::r_def3}. The honest block that triggers the fork race starts a new round, thus the round of the pool block of the race is $(0,0,\hQ_3)$ regardless of $\mathds{1}_{P_R}$, which gives \eqref{eq::GRP_rf}. With $\Lambda\equiv1$, Lemma~\ref{lem::rounds} gives $M_1=\phi_1(s_1)$, $M_s=\phi_1(s_1)e^{-s_2\hT}$ and $M_u=\phi_1(s_1)\phi_2(s_2)/(1-\theta)$, thus $\omega_3^x=0$, $\omega_2^1=0$, and by Lemma~\ref{lem::ratio}
    \begin{align*}
        \omega_2^{s}=\int_0^\infty\phi_1\,\hT e^{-s\hT}ds,\qquad \omega_2^{u}=\frac{1}{1-\theta}\int_0^\infty\phi_1\hat\phi_2\,ds.
    \end{align*}
    For $y\geq0$, the substitution $t=1+a_1s$ gives
    \begin{align*}
        \int_0^\infty\frac{ye^{-sy}}{1+a_1s}\,ds=\frac{y}{a_1}e^{y/a_1}E_1\Big(\frac{y}{a_1}\Big),
    \end{align*}
    which gives \eqref{eq::GRP_rs} with $y=\hT$. Since $\hat\phi_2(s)=(1-\theta)\E[\hQ_2e^{-s\hQ_2}]$ and $\hQ_2$ has the density $\frac{\mu_2e^{-\mu_2y}}{1-\theta}$ on $[0,\hT]$, the same identity gives \eqref{eq::GRP_ru}.
\end{remark}

\section{Reward Fractions for EXP}\label{app::exp_scheme}

\subsection{Model}
The rounds are defined by the canonical pool blocks as in (R1)--(R2) of Appendix~\ref{app::prop_scheme}, and the attack cycle evolves through the same cases. Under EXP scheme with decay rate $\lambda_d$, a share submitted $t$ time units before the end of its round receives the weight $w(t)=e^{-\lambda_dt}$, and the reward of a canonical pool block is split in proportion to the total weights of the shares of its round. As before, the pool submits blockwork at rate $\lambda_1a_i$ in mode $i$, of which $\lambda_1\alpha p_i$ is adversarial. For a round that closes at time $t_{\rm c}$, we define its \emph{score} $Z$ and \emph{adversarial score} $Z_A$ as
\begin{align}
    Z&=\int_{\text{round}}\lambda_1a(t)\,e^{-\lambda_d(t_{\rm c}-t)}\,dt,\label{eq::exp_score}\\
    Z_A&=\int_{\text{round}}\lambda_1\alpha p(t)\,e^{-\lambda_d(t_{\rm c}-t)}\,dt,\label{eq::exp_score_A}
\end{align}
where $a(t)=a_i$ and $p(t)=p_i$ if $t$ belongs to mode $i$, and we write $\mathbf{Z}=(Z_A,Z)$. The adversarial fraction of the reward of the round is $Z_A/Z$. We use
\begin{align}
    \kappa_1=\frac{\lambda_1}{\lambda_d},\ \kappa_2=\frac{\lambda_2}{\lambda_d},\ \chi_T=e^{-\lambda_dT},\ \mathbf{a}_i=(\alpha p_i,a_i),\label{eq::exp_kappa}
\end{align}
so that $\theta=\chi_T^{\kappa_2}$. As in Appendix~\ref{app::pplns_scheme}, the first phase and the fork race last $T_1,T_3\sim\mathrm{Exp}(\lambda_1)$, and a withholding phase that ends before the deadline (Case~4b, probability $1-\theta$) lasts $T_2'\sim(T_2\mid T_2<T)$ with $T_2\sim\mathrm{Exp}(\lambda_2)$; a withholding phase that reaches the deadline (Case~4a) lasts exactly $T$. We use the corresponding discount factors
\begin{align}
    \chi_1=e^{-\lambda_dT_1},\quad \chi_2'=e^{-\lambda_dT_2'},\quad \chi_3=e^{-\lambda_dT_3},\label{eq::exp_chi}
\end{align}
and $\chi_T$. Since $\mathbb{P}(\chi_1\leq u)=\mathbb{P}(T_1\geq-\ln u/\lambda_d)=u^{\kappa_1}$, $\chi_1$ and $\chi_3$ have the density $\kappa_1u^{\kappa_1-1}$ on $[0,1]$. Similarly, $\{T_2<T\}=\{e^{-\lambda_dT_2}>\chi_T\}$, thus $\chi_2'$ has the density
\begin{align}
    \frac{\kappa_2u^{\kappa_2-1}}{1-\theta}\,\mathds{1}_{\{u>\chi_T\}},\qquad u\in[0,1].\label{eq::exp_chi2_density}
\end{align}

\begin{lemma}[Score update]\label{lem::exp_update}
    If the pool mines in mode $i$ for a duration $t$ without closing the round, the score of the open round changes as
    \begin{align}
        \mathbf{Z}\ \mapsto\ e^{-\lambda_dt}\,\mathbf{Z}+\kappa_1\big(1-e^{-\lambda_dt}\big)\mathbf{a}_i.\label{eq::exp_update}
    \end{align}
\end{lemma}

\begin{Proof}
Moving the evaluation instant forward by $t$ multiplies the weight of every earlier share by $e^{-\lambda_dt}$, and the new shares contribute $\int_0^t\lambda_1a_ie^{-\lambda_d(t-t')}dt'=\kappa_1(1-e^{-\lambda_dt})a_i$ to $Z$ and $\kappa_1(1-e^{-\lambda_dt})\alpha p_i$ to $Z_A$.
\end{Proof}

Hence, the score updates over the first phase, the withholding phase ending before the deadline, the withholding phase ending at the deadline and the fork race are the affine maps
\begin{align}
    \mathcal{U}_1(\mathbf{z})&=\chi_1\mathbf{z}+\kappa_1(1-\chi_1)\mathbf{a}_1,\\
    \mathcal{U}_2(\mathbf{z})&=\chi_2'\mathbf{z}+\kappa_1(1-\chi_2')\mathbf{a}_2,\\
    \mathcal{U}_T(\mathbf{z})&=\chi_T\mathbf{z}+\kappa_1(1-\chi_T)\mathbf{a}_2,\\
    \mathcal{U}_3(\mathbf{z})&=\chi_3\mathbf{z}+\kappa_1(1-\chi_3)\mathbf{a}_3.
\end{align}

\subsection{The Carry Process}
Let $\mathbf{Z}^{(n)}$ be the \emph{carry}, i.e., the score of the open round evaluated at the start of cycle $n$. For a cycle that starts with carry $\mathbf{Z}^{(n)}=\mathbf{Z}$ and for a score $\mathbf{z}=(z_A,z)$, let $\varrho_d(\mathbf{z})=z_A/z$. Following the case-by-case analysis of Appendix~\ref{app::prop_scheme}, the canonical pool blocks are paid with the scores $\mathcal{U}_1(\mathbf{Z})$ in Case~3, $\mathcal{U}_T(\mathcal{U}_1(\mathbf{Z}))$ in Case~4a, $\mathcal{U}_2(\mathcal{U}_1(\mathbf{Z}))$ in Case~4b-ii and for the canonical fPoW of Case~4b-iii, and, in Case~4b-iii-B, $\mathcal{U}_3(\mathbf{0})$ if $\mathds{1}_{P_R}=1$ and $\mathcal{U}_3(\mathcal{U}_2(\mathcal{U}_1(\mathbf{Z})))$ if $\mathds{1}_{P_R}=0$. The carry left to the next cycle is
\begin{align}
    \mathbf{Z}^{(n+1)}=\begin{cases}
    \mathcal{U}_1(\mathbf{Z}), & \text{w.p. } \xi+\alpha q_1,\\
    \mathcal{U}_2(\mathcal{U}_1(\mathbf{Z})), & \text{w.p. } \alpha p_1\pi_A(1-\theta),\\
    \mathcal{U}_3(\mathcal{U}_2(\mathcal{U}_1(\mathbf{Z}))), & \text{w.p. } \varepsilon(1-\theta),\\
    \mathcal{U}_3(\mathbf{0}), & \text{w.p. } \eta_3,\\
    \mathbf{0}, & \text{w.p. } \eta_0,
    \end{cases}\label{eq::exp_carry}
\end{align}
with $\varepsilon,\eta_3,\eta_0$ in \eqref{eq::eps}--\eqref{eq::eta0}. That is, \eqref{eq::exp_carry} is \eqref{eq::carry_recursion} with the blockwork of each phase replaced by the corresponding score update under exponential decay. In particular, the reset cycles are the same, and by the renewal reward theorem $\rho_x=\E_\pi[B_x]/\E[B_C]$ with the carry in its stationary distribution $\pi$. The reward fractions are
\begin{align}
    r_1&=\E_\pi\big[\varrho_d\big(\mathcal{U}_1(\mathbf{Z})\big)\big],\\
    r_s&=\E_\pi\big[\varrho_d\big(\mathcal{U}_T(\mathcal{U}_1(\mathbf{Z}))\big)\big],\\
    r_u&=\E_\pi\big[\varrho_d\big(\mathcal{U}_2(\mathcal{U}_1(\mathbf{Z}))\big)\big],\\
    r_m&=\E_\pi\big[\varrho_d\big(\mathcal{U}_3(\mathcal{U}_2(\mathcal{U}_1(\mathbf{Z})))\big)\big],
\end{align}
where $\mathbf{Z}\sim\pi$ is independent of $\chi_1,\chi_2',\chi_3$, and $r_f=\mathds{1}_{P_R}a_3'+\overline{\mathds{1}_{P_R}}\,r_m$, since the fresh score $\mathcal{U}_3(\mathbf{0})=\kappa_1(1-\chi_3)\mathbf{a}_3$ has the fraction $a_3'$. 

\subsection{Transform of the Carry}
For a random score $\mathbf{Z}=(Z_A,Z)$, let
\begin{align}
    \mathcal{L}(s)=\E\big[e^{-sZ}\big],\quad \mathcal{L}_A(s)=\E\big[Z_Ae^{-sZ}\big],\quad s\geq0.\label{eq::exp_L}
\end{align}
For a mode $i$ and a measure $\mathfrak{m}$ on $[0,1]$, define the operator $\mathcal{G}^{\mathfrak{m}}_i(\mathcal{L},\mathcal{L}_A)=(\mathcal{L}',\mathcal{L}_A')$ by
\begin{align}
    \mathcal{L}'(s)&=\int_{[0,1]}e^{-s\kappa_1a_i(1-u)}\,\mathcal{L}(su)\,\mathfrak{m}(du),\label{eq::exp_G1}\\
    \mathcal{L}_A'(s)&=\int_{[0,1]}e^{-s\kappa_1a_i(1-u)}\Big(\kappa_1\alpha p_i(1-u)\,\mathcal{L}(su)\nonumber\\
    &\qquad\qquad+u\,\mathcal{L}_A(su)\Big)\,\mathfrak{m}(du),\label{eq::exp_G2}
\end{align}
and write $\mathcal{G}_1$ and $\mathcal{G}_3$ for the law of $\chi_1$ (and $\chi_3$), $\mathfrak{m}(du)=\kappa_1u^{\kappa_1-1}du$, with $i=1$ and $i=3$, $\mathcal{G}_2$ for the law \eqref{eq::exp_chi2_density} of $\chi_2'$ with $i=2$, and $\mathcal{G}_T$ for the point mass at $\chi_T$ with $i=2$.

\begin{lemma}\label{lem::exp_G}
    Let $\mathbf{Z}$ be a random score with transforms $(\mathcal{L},\mathcal{L}_A)$, and $\chi$ a random variable on $[0,1]$ with law $\mathfrak{m}$, independent of $\mathbf{Z}$. Then the transforms of $\chi\mathbf{Z}+\kappa_1(1-\chi)\mathbf{a}_i$ are $\mathcal{G}^{\mathfrak{m}}_i(\mathcal{L},\mathcal{L}_A)$.
\end{lemma}

\begin{Proof}
Conditioning on $\chi=u$, the new score is $(u Z_A+\kappa_1\alpha p_i(1-u),\,uZ+\kappa_1a_i(1-u))$, thus $\E[e^{-sZ'}\mid\chi=u]=e^{-s\kappa_1a_i(1-u)}\mathcal{L}(su)$ and $\E[Z_A'e^{-sZ'}\mid\chi=u]=e^{-s\kappa_1a_i(1-u)}\big(\kappa_1\alpha p_i(1-u)\mathcal{L}(su)+u\mathcal{L}_A(su)\big)$. Integrating against $\mathfrak{m}$ gives \eqref{eq::exp_G1}--\eqref{eq::exp_G2}.
\end{Proof}

\begin{theorem}\label{thm::exp}
    The transforms $(\mathcal{L},\mathcal{L}_A)$ of the stationary carry are the unique bounded solution of
    \begin{align}
        (\mathcal{L},\mathcal{L}_A)=&\ \eta_0\,(1,0)+\eta_3\,\mathcal{G}_3(1,0)\nonumber\\
        &+(\xi+\alpha q_1)\,\mathcal{G}_1(\mathcal{L},\mathcal{L}_A)\nonumber\\
        &+\alpha p_1\pi_A(1-\theta)\,\mathcal{G}_2\mathcal{G}_1(\mathcal{L},\mathcal{L}_A)\nonumber\\
        &+\varepsilon(1-\theta)\,\mathcal{G}_3\mathcal{G}_2\mathcal{G}_1(\mathcal{L},\mathcal{L}_A),\label{eq::exp_fixed}
    \end{align}
    where $(1,0)$ is the pair of transforms of the zero carry. With
    \begin{align}
        (\mathcal{L}^{1},\mathcal{L}_A^{1})&=\mathcal{G}_1(\mathcal{L},\mathcal{L}_A),\\
        (\mathcal{L}^{s},\mathcal{L}_A^{s})&=\mathcal{G}_T\mathcal{G}_1(\mathcal{L},\mathcal{L}_A),\\
        (\mathcal{L}^{u},\mathcal{L}_A^{u})&=\mathcal{G}_2\mathcal{G}_1(\mathcal{L},\mathcal{L}_A),\\
        (\mathcal{L}^{m},\mathcal{L}_A^{m})&=\mathcal{G}_3\mathcal{G}_2\mathcal{G}_1(\mathcal{L},\mathcal{L}_A),
    \end{align}
    the expected adversarial reward fractions under EXP scheme are
    \begin{align}
        r_x&=\int_0^\infty\mathcal{L}_A^{x}(s)\,ds,\qquad x\in\{1,s,u,m\},\\
        r_f&=\mathds{1}_{P_R}\,a_3'+\overline{\mathds{1}_{P_R}}\,r_m.
    \end{align}
\end{theorem}

\begin{Proof}
Taking the transforms \eqref{eq::exp_L} of both sides of \eqref{eq::exp_carry} in stationarity, and applying Lemma~\ref{lem::exp_G} once per phase, in the order in which the phases occur, gives \eqref{eq::exp_fixed}. The first component of \eqref{eq::exp_fixed} does not involve $\mathcal{L}_A$, and the second is affine in $\mathcal{L}_A$ for a given $\mathcal{L}$. Since $e^{-s\kappa_1a_i(1-u)}\leq1$ and $u\leq1$, both are contractions in the supremum norm with factor $\xi+\alpha q_1+(1-\theta)(\alpha p_1\pi_A+\varepsilon)=1-\eta_0-\eta_3<1$, hence the solution is unique and the iteration of \eqref{eq::exp_fixed} converges to it. By Lemma~\ref{lem::exp_G}, $(\mathcal{L}^{x},\mathcal{L}_A^{x})$ are the transforms of the paid scores. Finally, for a paid score $\mathbf{Z}$, use $\E[\varrho_d(\mathbf{Z})]=\int_0^\infty\E[Z_Ae^{-sZ}]ds$ by $\frac{1}{z}=\int_0^\infty e^{-sz}ds$ and Tonelli's theorem to get the reward fractions.
\end{Proof}

Every operator in \eqref{eq::exp_fixed} is a one-dimensional integral over $u\in[0,1]$ that evaluates $\mathcal{L}$ and $\mathcal{L}_A$ at $su\leq s$, thus \eqref{eq::exp_fixed} can be solved on a grid in $s$ by iteration, and $r_x$ follows by one-dimensional quadrature.

\section{Revenue Ratios of Pool Hopping}\label{app::hop}
\begin{lemma}\label{lem::hop_round}
    Under PAH, the rounds are i.i.d. Let $\tau$, $y(t)$ in \eqref{eq::hop_age} and $Y_\tau=y(\tau)$ be as defined in Section~\ref{sec::hop}. Then,
    \begin{itemize}[leftmargin=*]
        \item[(i)] $\mathbb{P}(\tau>t)=\bar F_\tau(t)$ with $\bar F_\tau(t)=e^{-\lambda_1a_yt}$ for $t<t^*$ and $\bar F_\tau(t)=e^*e^{-\lambda_1a_o(t-t^*)}$ for $t\geq t^*$, and the pool block is found by the adversary with probability $a'(\tau)$;
        \item[(ii)] the expected numbers of honest blocks outside the pool and solo adversarial blocks found before $\tau$ are $n_R$ and $n_S$;
        \item[(iii)] $Y_\tau\sim\mathrm{Exp}(1)$ for every policy $(p_y,p_o,t^*)$.
    \end{itemize}
\end{lemma}

\begin{Proof}
A round starts at a stopping time of the mining processes and the policy depends only on the time since the start of the round, hence the rounds are i.i.d.~by the memoryless property. At time $t$, the pool mines with the hash fraction $a(t)$, hence the pool blocks arrive with the intensity $\lambda_1a(t)$, which gives $\bar F_\tau$. A pool block at time $t$ is adversarial with probability $\frac{\alpha p(t)}{a(t)}=a'(t)$. The outside and solo blocks arrive with the intensities $\lambda_1\xi$ and $\lambda_1\alpha q(t)$ and do not end the round, thus their expected numbers are $\int_0^\infty\bar F_\tau(t)\lambda_1\xi\,dt$ and $\int_0^\infty\bar F_\tau(t)\lambda_1\alpha q(t)\,dt$, which give \eqref{eq::hop_nr} and \eqref{eq::hop_ns}.

For (iii), the idea is to measure the round by the work done in it instead of by time. Every hash is a valid block with the same probability, regardless of who computes it and how fast. Hence, the probability that the pool finds a block depends only on how much blockwork the pool has done, and not on how long it took. The blockwork age counts exactly this work, thus every additional unit of blockwork age carries the same chance of ending the round, i.e., in units of blockwork age the round ends at the constant rate $1$. The power allocation of the adversary only changes how fast the blockwork age grows, i.e., faster with $p_y$ and slower with $p_o$, but not how much blockwork age a round lasts.

Formally, \eqref{eq::hop_age} is deterministic, continuous and strictly increasing since $a(t)\geq\beta>0$, and $y(t^*)=y^*$. By (i),
\begin{align}
    \mathbb{P}(\tau>t)=\exp\Big(-\int_0^t\lambda_1a(t')\,dt'\Big)=e^{-y(t)},\label{eq::hop_tau_age}
\end{align}
and since $y(\cdot)$ is strictly increasing, $\{\tau>t\}=\{Y_\tau>y(t)\}$. Hence, $\mathbb{P}(Y_\tau>y)=e^{-y}$ for all $y\geq0$, i.e., $Y_\tau\sim\mathrm{Exp}(1)$. Equivalently, the probability of a pool block in $[t,t+dt)$ is $\lambda_1a(t)\,dt=dy(t)$, the increment of the blockwork age, hence the pool blocks form a Poisson process of unit rate in units of blockwork age, which is the time change of a Poisson process.
\end{Proof}
 
\subsection{Proof of Theorem~\ref{thm::hop}}
Nothing is withheld, therefore every block is canonical and $\delta^{\rm PAH}=1$. Every round contains exactly one pool block, which pays $1$ in total, and by Lemma~\ref{lem::hop_round}(ii), $n_S$ solo and $n_R$ outside blocks on average, each paying $1$. By the renewal reward theorem, $\rho_x$ is the expected reward of $x$ per round divided by the expected number of canonical blocks per round, $n_S+n_R+1$, which gives \eqref{eq::hop_rho_A}-\eqref{eq::hop_rho_R} with $r_h$ the expected adversarial reward fraction of the pool block. It remains to compute $r_h$.
 
\emph{PROP:} The pool block pays the adversarial fraction of the blockwork of the round. By Lemma~\ref{lem::hop_round}(iii), $Y_\tau\sim\mathrm{Exp}(1)$, and the adversarial blockwork at blockwork age $y$ is $b_A(y)=a_y'y$ for $y<y^*$ and $b_A(y)=a_o'y+(a_y'-a_o')y^*$ for $y\geq y^*$. Hence,
\begin{align}
    r_h^{\rm PROP}&=\E\bigg[\frac{b_A(Y_\tau)}{Y_\tau}\bigg]=\int_0^{y^*}e^{-y}a_y'\,dy\nonumber\\
    &\quad+\int_{y^*}^\infty e^{-y}\Big(a_o'+\frac{(a_y'-a_o')y^*}{y}\Big)dy,\label{eq::hop_rprop_proof}
\end{align}
which is \eqref{eq::hop_r_prop} since $\int_{y^*}^\infty\frac{e^{-y}}{y}dy=E_1(y^*)$.
 
\emph{EXP:} With $a(t)$ and $p(t)$ given by the policy \eqref{eq::hop_policy}, the score and the adversarial score \eqref{eq::exp_score}--\eqref{eq::exp_score_A} of the open round evaluated at time $t$ are deterministic:
\begin{align}
    Z(t)=&\int_0^t\lambda_1a(t')e^{-\lambda_d(t-t')}dt',\\
    Z_A(t)=&\int_0^t\lambda_1\alpha p(t')e^{-\lambda_d(t-t')}dt',
\end{align}
and the pool block at time $\tau$ pays the adversary $Z_A(\tau)/Z(\tau)$. For $t<t^*$, the pool hashing is constant, thus $Z_A/Z=a_y'$, and the pool block falls before $t^*$ with probability $1-e^*$. For $t=t^*+u$ with $u\geq0$, by Lemma~\ref{lem::exp_update},
\begin{align}
    \frac{Z(t)}{\kappa_1}&=a_y(1-\chi^*)e^{-\lambda_du}+a_o\big(1-e^{-\lambda_du}\big),\\
    \frac{Z_A(t)}{\kappa_1}&=\alpha p_y(1-\chi^*)e^{-\lambda_du}+\alpha p_o\big(1-e^{-\lambda_du}\big),
\end{align}
and, given $\tau\geq t^*$, $\tau-t^*\sim\mathrm{Exp}(\lambda_1a_o)$ by Lemma~\ref{lem::hop_round}(i). Hence,
\begin{align}
    r_h^{\rm EXP}=&\ a_y'(1-e^*)\nonumber\\
    &+e^*\int_0^\infty\lambda_1a_oe^{-\lambda_1a_ou}\,\frac{Z_A(t^*+u)}{Z(t^*+u)}\,du,
\end{align}
and the substitution $v=e^{-\lambda_du}$, under which $\lambda_1a_oe^{-\lambda_1a_ou}du=-\kappa_ov^{\kappa_o-1}dv$, gives \eqref{eq::hop_r_exp}.\hfill$\blacksquare$

\subsection{All-or-Nothing Hopping}
Let $p_y=1$ and $p_o=0$, i.e., the adversary mines only in the pool during the first $t^*$ time units of a round, which correspond to the first $y^*=\lambda_1(\alpha+\beta)t^*$ blockworks, and only solo afterwards. Then, $a_y=\alpha+\beta$, $a_o=\beta$, $a_y'=\frac{\alpha}{\alpha+\beta}$, $a_o'=0$, and
\begin{align}
    n_S=\frac{\alpha}{\beta}e^*,\qquad n_S+n_R+1=\frac{1-e^*}{\alpha+\beta}+\frac{e^*}{\beta}.\label{eq::hop_aon_n}
\end{align}

\begin{corollary}\label{cor::hop_prop}
    Let $\beta'=\frac{\beta}{\alpha+\beta}$. Under round-based PROP,
    \begin{align}
        \frac{\rho_A}{\alpha}-1=\frac{\beta'\,y^*E_1(y^*)}{\beta'+(1-\beta')e^*}.\label{eq::hop_RER}
    \end{align}
    Hence, in this case, the relative extra reward depends on only $\alpha/\beta$, and it is maximized at the root $y^*$ of
    \begin{align}
        &\big(E_1(y^*)-e^{-y^*}\big)\big(\beta'+(1-\beta')e^{-y^*}\big)\nonumber\\
        &\qquad+(1-\beta')\,e^{-y^*}y^*E_1(y^*)=0.\label{eq::hop_opt}
    \end{align}
    As $\alpha/\beta\to0$, i.e., $\beta'\to1$, \eqref{eq::hop_RER} tends to $y^*E_1(y^*)$, which is maximized at the root $y^*\approx0.4348$ of $e^{y^*}E_1(y^*)=1$, where it equals $y^*e^{-y^*}\approx0.2815$, which recovers the classical pool hopping result of~\cite{raulo2011optimal,rosenfeld2011analysisbitcoinpooledmining} for a negligible miner. For a non-negligible adversary, its own shares dilute the young rounds it mines in, which is captured by $\beta'$ in \eqref{eq::hop_RER}; as a result, the gain decreases and the optimal threshold $y^*$ increases with $\alpha/\beta$.
\end{corollary}

\begin{Proof}
By \eqref{eq::hop_r_prop}, $r_h^{\rm PROP}=\frac{\alpha}{\alpha+\beta}\big(1-e^*+y^*E_1(y^*)\big)$, hence by \eqref{eq::hop_rho_A} and \eqref{eq::hop_aon_n},
\begin{align}
    \frac{\rho_A}{\alpha}&=\frac{\frac{e^*}{\beta}+\frac{1-e^*+y^*E_1(y^*)}{\alpha+\beta}}{\frac{e^*}{\beta}+\frac{1-e^*}{\alpha+\beta}}\\
    &=1+\frac{y^*E_1(y^*)/(\alpha+\beta)}{e^*/\beta+(1-e^*)/(\alpha+\beta)},
\end{align}
and multiplying the numerator and the denominator of the last fraction by $\beta$ gives \eqref{eq::hop_RER}. The derivative of $yE_1(y)$ is $E_1(y)-e^{-y}$ and that of $\beta'+(1-\beta')e^{-y}$ is $-(1-\beta')e^{-y}$, which gives \eqref{eq::hop_opt}. For $\beta'=1$, \eqref{eq::hop_opt} reduces to $E_1(y^*)=e^{-y^*}$.
\end{Proof}

\begin{corollary}\label{cor::hop_exp}
    Let $\varsigma=1-\frac{\alpha+\beta}{\beta}(1-\chi^*)$. Under EXP, with $\kappa_o=\lambda_1\beta/\lambda_d$,
    \begin{align}
        r_h^{\rm EXP}=&\ \frac{\alpha}{\alpha+\beta}(1-e^*)+e^*\,\frac{\kappa_o}{\kappa_o+1}\,\frac{\alpha(1-\chi^*)}{\beta}\nonumber\\
        &\times{}_2F_1(1,\kappa_o+1;\kappa_o+2;\varsigma),\label{eq::hop_r_exp_aon}
    \end{align}
    where ${}_2F_1$ is the Gaussian hypergeometric function.
\end{corollary}

\begin{Proof}
    With $p_o=0$ the integrand of \eqref{eq::hop_r_exp} is $v^{\kappa_o-1}\cdot\frac{\alpha(1-\chi^*)v}{\beta(1-\varsigma v)}$, and $\int_0^1\frac{v^{\kappa_o}}{1-\varsigma v}dv=\frac{1}{\kappa_o+1}\,{}_2F_1(1,\kappa_o+1;\kappa_o+2;\varsigma)$ by Euler's integral representation; note that $\varsigma<1$ for $t^*>0$.
\end{Proof}

Unlike \eqref{eq::hop_RER}, the gain under EXP depends on $\alpha$ and $\beta$ separately, since the decay acts in time while the rounds end at the rate $\lambda_1a(t)$, which depends on the size of the pool. If the rounds are long compared to $1/\lambda_d$, the score of an old round saturates, its composition is that of the recent shares only, and hopping gains little; if the rounds are short compared to $1/\lambda_d$, EXP behaves like PROP.

\subsection{Pool Hopping Under $W$-PPLNS}
The following is the hopping-proofness of PPLNS shown in~\cite[Sec.~3.3]{rosenfeld2011analysisbitcoinpooledmining}, which we state for an adversary of any size and any power allocation policy. 

\begin{proposition}[Rosenfeld \cite{rosenfeld2011analysisbitcoinpooledmining}]\label{prop::hop_pplns}
    Under $W$-PPLNS with constant block rewards, every policy under which the adversary submits all its shares immediately and chooses its power allocation based only on the past of the mining processes gives $\rho_A=\alpha$.
\end{proposition}

\begin{Proof}
Consider a share of blockwork $dv$. It receives the fraction $dv/W$ of the reward of every pool block found while it is among the last $W$ blockworks of pool shares, i.e., during the next $W$ blockworks of pool work. Measured by the pool blockwork, the pool blocks arrive as a Poisson process of unit rate whatever the allocation is, by the time change argument in the proof of Lemma~\ref{lem::hop_round}(iii), and its future is independent of the past. Hence, the expected number of pool blocks during the next $W$ blockworks is $W$, given anything that is known when the share is submitted, and the expected reward of the share is $dv$, which is the expected reward of the blockwork $dv$ mined solo. Summing over all shares, the expected reward of the adversary per unit time equals that of mining solo with all its hashpower, which is $\alpha$ times the expected canonical reward per unit time, since all blocks are canonical.  
\end{Proof}
Thus, a gain under $W$-PPLNS requires information about future pool blocks that is not contained in the past of the mining processes, such as a withheld fPoW whose release time the adversary controls, which is the idea behind lie-in-wait~\cite{rosenfeld2011analysisbitcoinpooledmining} and T-PAW~\cite{doger2026temporarypoweradjustingwithholding}. In other words, under $W$-PPLNS, any extra reward of PAW/T-PAW over honest mining stems from withholding and not from power adjusting alone.

\begin{table}[t]
    \centering
    \setlength{\tabcolsep}{4pt}
    \resizebox{\columnwidth}{!}{%
    \begin{tabular}{ll|cc|cc|cc}
    \toprule
     & & & & \multicolumn{2}{c|}{RER (\%)} & \multicolumn{2}{c}{$\delta$}\\
    Attack & Scheme & $\beta$ & $\gamma$ & Theory & MC & Theory & MC\\
    \midrule
    T-PAW & PPLNS & 0.05 & 0 & 0.083 & 0.062 $\pm$ 0.092 & 1.0014 & 1.0015 $\pm$ 0.0002\\
    T-PAW & PPLNS & 0.05 & 0.5 & 0.195 & 0.167 $\pm$ 0.066 & 1.0032 & 1.0032 $\pm$ 0.0002\\
    T-PAW & PPLNS & 0.1 & 0 & 0.177 & 0.111 $\pm$ 0.046 & 1.0031 & 1.0031 $\pm$ 0.0002\\
    T-PAW & PPLNS & 0.1 & 0.5 & 0.376 & 0.375 $\pm$ 0.071 & 1.0062 & 1.0062 $\pm$ 0.0002\\
    T-PAW & PROP & 0.05 & 0 & 0.194 & 0.157 $\pm$ 0.085 & 1.0016 & 1.0015 $\pm$ 0.0003\\
    T-PAW & PROP & 0.05 & 0.5 & 0.544 & 0.511 $\pm$ 0.077 & 1.0046 & 1.0046 $\pm$ 0.0002\\
    T-PAW & PROP & 0.1 & 0 & 0.298 & 0.295 $\pm$ 0.080 & 1.0028 & 1.0029 $\pm$ 0.0002\\
    T-PAW & PROP & 0.1 & 0.5 & 0.724 & 0.667 $\pm$ 0.071 & 1.0075 & 1.0074 $\pm$ 0.0002\\
    T-PAW & EXP & 0.05 & 0 & 2.382 & 2.434 $\pm$ 0.053 & 1.0031 & 1.0032 $\pm$ 0.0002\\
    T-PAW & EXP & 0.05 & 0.5 & 3.425 & 3.403 $\pm$ 0.063 & 1.0054 & 1.0054 $\pm$ 0.0002\\
    T-PAW & EXP & 0.1 & 0 & 1.955 & 1.961 $\pm$ 0.078 & 1.0043 & 1.0043 $\pm$ 0.0002\\
    T-PAW & EXP & 0.1 & 0.5 & 2.800 & 2.805 $\pm$ 0.084 & 1.0072 & 1.0071 $\pm$ 0.0003\\
    \midrule
    PAH & PROP & 0.05 & -- & 17.603 & 17.601 $\pm$ 0.009 & 1.0000 & 1.0000 $\pm$ 0.0000\\
    PAH & PROP & 0.1 & -- & 21.496 & 21.500 $\pm$ 0.007 & 1.0000 & 1.0000 $\pm$ 0.0000\\
    PAH & EXP & 0.05 & -- & 1.523 & 1.524 $\pm$ 0.004 & 1.0000 & 1.0000 $\pm$ 0.0000\\
    PAH & EXP & 0.1 & -- & 3.252 & 3.253 $\pm$ 0.005 & 1.0000 & 1.0000 $\pm$ 0.0000\\
    \bottomrule
    \end{tabular}}
    \caption{Theory and Monte Carlo (MC) simulation of the relative extra reward and the block redundancy ratio at the optimal attack parameters for $\alpha=0.05$, with $W=2$ for PPLNS and $\lambda_d=2\lambda_1$ for EXP. MC results are the mean and the 99\% confidence half-width over 20 independent runs of $10^{7}$ attack cycles (T-PAW) or rounds (PAH).}
    \label{tab::validation}
\end{table}

\begin{table*}[t]
    \centering
    \setlength{\tabcolsep}{4pt}
    \resizebox{2\columnwidth}{!}{%
    \begin{tabular}{ll|cc|cc|cc|cc|cc}
    \toprule
     & & & & \multicolumn{2}{c|}{$r_1$ ($r_h$)} & \multicolumn{2}{c|}{$r_s$} & \multicolumn{2}{c|}{$r_u$} & \multicolumn{2}{c}{$r_f$} \\
    Attack & Scheme & $\beta$ & $\gamma$ & Theory & MC & Theory & MC & Theory & MC & Theory & MC\\
    \midrule
    T-PAW & PPLNS & 0.05 & 0 & 0.0323 & 0.0323 $\pm$ 0.0000 & 0.0795 & 0.0795 $\pm$ 0.0001 & 0.0488 & 0.0486 $\pm$ 0.0002 & 0.0480 & 0.0478 $\pm$ 0.0003\\
    T-PAW & PPLNS & 0.05 & 0.5 & 0.0615 & 0.0615 $\pm$ 0.0000 & 0.1403 & 0.1404 $\pm$ 0.0002 & 0.0819 & 0.0818 $\pm$ 0.0001 & 0.0803 & 0.0801 $\pm$ 0.0003\\
    T-PAW & PPLNS & 0.1 & 0 & 0.0334 & 0.0334 $\pm$ 0.0000 & 0.0821 & 0.0821 $\pm$ 0.0001 & 0.0499 & 0.0499 $\pm$ 0.0001 & 0.0482 & 0.0482 $\pm$ 0.0002\\
    T-PAW & PPLNS & 0.1 & 0.5 & 0.0591 & 0.0591 $\pm$ 0.0000 & 0.1365 & 0.1364 $\pm$ 0.0001 & 0.0785 & 0.0785 $\pm$ 0.0001 & 0.0755 & 0.0755 $\pm$ 0.0002\\
    T-PAW & PROP & 0.05 & 0 & 0.0607 & 0.0607 $\pm$ 0.0000 & 0.1288 & 0.1289 $\pm$ 0.0004 & 0.0977 & 0.0975 $\pm$ 0.0008 & 0.5000 & 0.5000 $\pm$ 0.0000\\
    T-PAW & PROP & 0.05 & 0.5 & 0.0989 & 0.0989 $\pm$ 0.0000 & 0.2144 & 0.2148 $\pm$ 0.0006 & 0.1552 & 0.1554 $\pm$ 0.0003 & 0.5000 & 0.5000 $\pm$ 0.0000\\
    T-PAW & PROP & 0.1 & 0 & 0.0525 & 0.0524 $\pm$ 0.0000 & 0.1109 & 0.1110 $\pm$ 0.0002 & 0.0849 & 0.0849 $\pm$ 0.0004 & 0.3333 & 0.3333 $\pm$ 0.0000\\
    T-PAW & PROP & 0.1 & 0.5 & 0.0786 & 0.0786 $\pm$ 0.0000 & 0.1735 & 0.1737 $\pm$ 0.0003 & 0.1260 & 0.1260 $\pm$ 0.0001 & 0.3333 & 0.3333 $\pm$ 0.0000\\
    T-PAW & EXP & 0.05 & 0 & 0.1750 & 0.1750 $\pm$ 0.0000 & 0.3798 & 0.3798 $\pm$ 0.0000 & 0.2943 & 0.2945 $\pm$ 0.0006 & 0.5000 & 0.5000 $\pm$ 0.0000\\
    T-PAW & EXP & 0.05 & 0.5 & 0.2087 & 0.2087 $\pm$ 0.0000 & 0.4271 & 0.4271 $\pm$ 0.0000 & 0.3404 & 0.3402 $\pm$ 0.0003 & 0.5000 & 0.5000 $\pm$ 0.0000\\
    T-PAW & EXP & 0.1 & 0 & 0.1135 & 0.1135 $\pm$ 0.0000 & 0.2553 & 0.2553 $\pm$ 0.0000 & 0.1951 & 0.1951 $\pm$ 0.0003 & 0.3333 & 0.3333 $\pm$ 0.0000\\
    T-PAW & EXP & 0.1 & 0.5 & 0.1351 & 0.1351 $\pm$ 0.0000 & 0.2877 & 0.2877 $\pm$ 0.0000 & 0.2260 & 0.2260 $\pm$ 0.0002 & 0.3333 & 0.3333 $\pm$ 0.0000\\
    \midrule
    PAH & PROP & 0.05 & -- & 0.3630 & 0.3630 $\pm$ 0.0000 & -- & -- & -- & -- & -- & --\\
    PAH & PROP & 0.1 & -- & 0.2298 & 0.2298 $\pm$ 0.0000 & -- & -- & -- & -- & -- & --\\
    PAH & EXP & 0.05 & -- & 0.0808 & 0.0808 $\pm$ 0.0000 & -- & -- & -- & -- & -- & --\\
    PAH & EXP & 0.1 & -- & 0.0678 & 0.0678 $\pm$ 0.0000 & -- & -- & -- & -- & -- & --\\
    \bottomrule
    \end{tabular}}
    \caption{Theory and Monte Carlo (MC) simulation of the reward fractions at the optimal attack parameters, with the same setting as Table~\ref{tab::validation}. For PAH, the first column gives the expected adversarial fraction $r_h$ of a pool block. A dash marks an event that does not occur at the optimum.}
    \label{tab::validation_r}
\end{table*}

\section{Monte Carlo Verifications}\label{app::mc}
To verify the analytical expressions, we simulate the attacks directly, without using any of the derivations in Appendices~\ref{app::pplns_scheme}--\ref{app::hop}. For T-PAW, we simulate many independent chains of attack cycles in parallel. Each chain follows the cases of \figref{fig::cycle_events} and keeps the state of its pool explicitly: the last $W$ blockworks of pool shares, as a list of segments, for PPLNS; the blockwork of the open round for PROP; and the score of the open round, updated by Lemma~\ref{lem::exp_update}, for EXP. Every canonical pool block pays the adversary according to the current state. For PROP and EXP, we discard a burn-in of $\lceil5/\beta\rceil+10$ cycles so that the carry is stationary, and for PPLNS, we discard the cycles until the initial window is replaced. For PAH, the rounds are i.i.d.~by Lemma~\ref{lem::hop_round}, hence we simulate them independently. Since the numbers of outside and solo blocks in a round do not affect the following rounds, we replace them with their conditional means given the round length, which keeps the estimator unbiased and reduces its variance. For each case, we use the optimal parameters of Section~\ref{sec::numerical} for $\alpha=0.05$.

Table~\ref{tab::validation} shows that the simulated RER and $\delta$ agree with the theory. For PAH, the confidence intervals are below $0.01\%$, and the theory and the simulation agree to the third digit, e.g., $17.603\%$ versus $17.601\%\pm0.009\%$ under PROP for $\beta=0.05$. For T-PAW, the confidence intervals are wider, since the RER is a small difference between two revenue ratios, $\rho_A/\alpha-1$, and the reward of a cycle has a large variance. The theory lies within the $99\%$ confidence interval in all cases except PPLNS with $(\beta,\gamma)=(0.1,0)$, where it lies just outside.  Both this deviation as well as the fact that the T-PAW gain is of the same order as the confidence interval for PPLNS and PROP, imply  these RER values alone are not a sharp test.

A sharper test is given by the reward fractions in Table~\ref{tab::validation_r}. The revenue ratios of T-PAW depend on the payout scheme only through $r_1,r_s,r_u,r_f$, hence the scheme-dependent part of the analysis can be verified through these fractions alone. They are averages over many pool blocks, thus their confidence intervals are much smaller. The theory and the simulation agree to the fourth digit for every scheme, including the effect of the carry under PROP and EXP, e.g., $r_u=0.1552$ versus $0.1554\pm0.0003$ under PROP for $(\beta,\gamma)=(0.05,0.5)$. Under PPLNS, the simulated $r_u$ and $r_f$ are slightly below the theory, by at most $2\times10^{-4}$. At the optimum, $p_3=1$, hence the pool block of the fork race is paid $r_f=a_3'$ exactly under PROP and EXP, since it closes a fresh round. Finally, the simulated $r_h$ of PAH matches the theory in every case, which verifies Theorem~\ref{thm::hop} and the blockwork age argument of Lemma~\ref{lem::hop_round}.

\bibliographystyle{IEEEtran}
\bibliography{blockchain}
\end{document}